\documentclass[11pt]{article}

\usepackage{amsfonts}

\usepackage{amssymb}
\usepackage{latexsym}
\usepackage{graphicx}
\usepackage[english]{babel}
\usepackage[font={small}]{caption}

\usepackage{amsfonts}
\usepackage{latexsym}
\usepackage{graphicx}
\usepackage[english]{babel}
\begin{document}
\input epsf

\def\p{\partial}
\def\h{{1\over 2}}
\def\be{\begin{equation}}
\def\bea{\begin{eqnarray}}
\def\ee{\end{equation}}
\def\eea{\end{eqnarray}}
\def\d{\partial}
\def\la{\lambda}
\def\eps{\epsilon}
\def\bb{\bigskip}
\def\mm{\medskip}
\newcommand{\dm}{\begin{displaymath}}
\newcommand{\edm}{\end{displaymath}}
\renewcommand{\b}{\tilde{B}}
\newcommand{\gm}{\Gamma}
\newcommand{\ac}[2]{\ensuremath{\{ #1, #2 \}}}
\renewcommand{\ell}{l}
\newcommand{\z}{\ell}
\newcommand{\newsection}[1]{\section{#1} \setcounter{equation}{0}}
\def\bb{$\bullet$}
\def\Qbar{{\bar Q}_1}
\def\QPbar{{\bar Q}_p}

\def\q{\quad}

\def\bn{B_\circ}

\let\a=\alpha \let\b=\beta \let\g=\gamma \let\d=\delta \let\e=\epsilon
\let\c=\chi \let\th=\theta  \let\k=\kappa
\let\l=\lambda \let\m=\mu \let\n=\nu \let\x=\xi \let\r=\rho
\let\s=\sigma \let\t=\tau
\let\vp=\varphi \let\vep=\varepsilon
\let\w=\omega      \let\G=\Gamma \let\D=\Delta \let\Th=\Theta
                     \let\P=\Pi \let\S=\Sigma

\def\h{{1\over 2}}
\def\t{\tilde}
\def\r{\rightarrow}
\def\nn{\nonumber\\}
\let\bm=\bibitem
\def\Kt{{\tilde K}}
\def\b{\bigskip}

\let\p=\partial

\begin{flushright}
\end{flushright}
\vspace{20mm}
\begin{center}
{\LARGE  A model with no firewall}
\\
\vspace{18mm}
{\bf  Samir D. Mathur }\\

\vspace{8mm}
Department of Physics,\\ The Ohio State University,\\ Columbus,
OH 43210, USA\\mathur.16@osu.edu\\
\vspace{4mm}

\end{center}
\vspace{10mm}
\thispagestyle{empty}
\begin{abstract}

\b

\b

We construct a model which illustrates the conjecture of fuzzball complementarity. In the fuzzball paradigm, the black hole microstates have no interior, and radiate unitarily from their surface through quanta of energy $E\sim T$. But  quanta with $E\gg T$ impinging on the fuzzball create large collective excitations of the fuzzball surface. The dynamics of such excitations must be studied as an evolution in superspace, the space of all fuzzball solution $|F_i\rangle$. The states in this superspace are arranged in a hierarchy of `complexity'. We argue that   evolution towards higher complexity  maps, through a duality  analogous to AdS/CFT, to infall inside the horizon of the traditional hole.  We explain how the large degeneracy of fuzzball states leads to a breakdown of the principle of equivalence at the threshold of horizon formation. We recall that the firewall argument did not invoke  the limit  $E\gg T$  when considering a complementary picture; on the contrary it focused on the dynamics of the $E\sim T$ modes which contribute to Hawking radiation. This loophole allows the dual description conjectured in fuzzball complementarity.

\end{abstract}
\vskip 1.0 true in

\newpage
\setcounter{page}{1}

\section{Introduction}

The `no-hair' theorems for black holes imply that the region around the horizon must be a vacuum \cite{nohair}. Hawking showed that entangled pairs are created at such a horizon, and this  creates a conflict with quantum unitarity near the end of evaporation \cite{hawking}. It has been shown (using strong subadditivity of quantum entanglement entropy)  that Hawking's argument is stable to small corrections; thus one needs a change of {\it order unity } in the evolution of low energy modes at the horizon \cite{cern}.

\begin{figure}[htbp]
   \begin{center}
   \includegraphics[scale=.8]{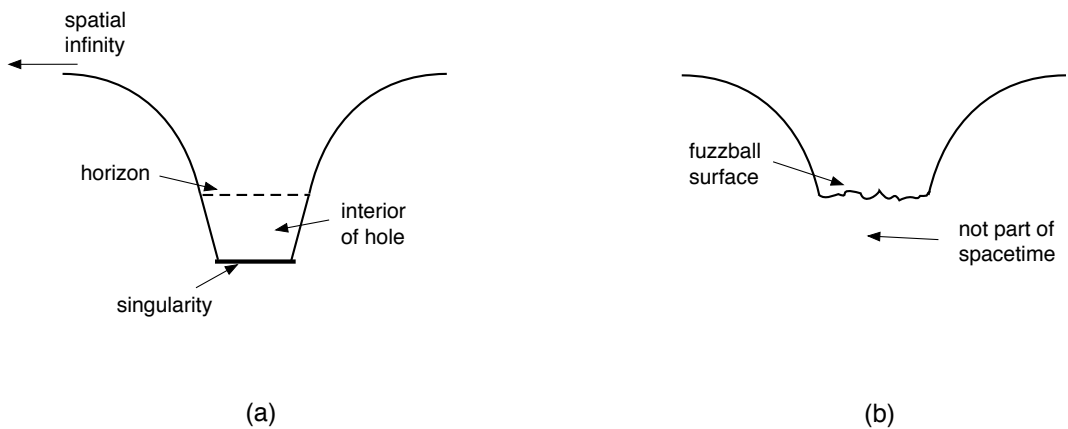}
   \caption{(a) The conventional picture of a black hole.  (b) the fuzzball picture: spacetime ends just outside the horizon in a quantum mess. }
   \label{fuzzball}
   \end{center}
   \end{figure}

In string theory we indeed find a complete alteration of the horizon; black hole microstates are fuzzballs, where spacetime ends just outside the expected horizon radius (fig.\ref{fuzzball}). The fuzzball radiates from its surface just like a piece of coal: the {\it rate} of radiation agrees with the Hawking rate \cite{myers,cm1}, but the entanglement structure is completely different.  The large modification away from the semiclassical hole is explained by the largeness of the Bekenstein entropy: there is a small probability for a collapsing star to tunnel into a fuzzball microstate, but this smallness is cancelled the the large number  ${\cal N} \sim Exp[S_{bek}]$   of fuzzballs that the star can transition to \cite{tunnel}. This resolves the information paradox.

But this leaves a last question: is there any role at all for the traditional black hole metric which {\it did} have a region interior to the horizon? In \cite{fuzzcomp} the conjecture of fuzzball complementarity was developed to address this question. According to this conjecture, there is no complementary description for quanta at energies $E\sim T$. Such quanta carry out the information of the hole and must therefore differ in their details between different fuzzballs; thus their physics cannot be replaced by the physics of an `information free' vacuum region.  But  now consider infalling quanta with energy $E\gg T$. The impact of such quanta creates a large deformation of the fuzzball. At leading order in $T/E$, this deformation is independent of the precise choice of fuzzball microstate. The conjecture states that the evolution of these large deformations can be mapped to the evolution of radially infalling modes in the {\it traditional} black hole interior. Since $T$ is very low for a large black hole, this conjecture would give a good map to the traditional hole for typical infalling objects. Note that this complementary mapping is in the nature of a duality. The infalling object does not in any way `go through' the horizon; what happens instead is that the spectrum of excitations of the fuzzball surface agrees, to a good aproximation, with the spectrum of infalling modes in the traditional black hole.  

In this paper we will give an explicit model for how such a complementarity can work. We proceed as follows:

\b

\begin{figure}[h]
\begin{center}
 \includegraphics[scale=.82] {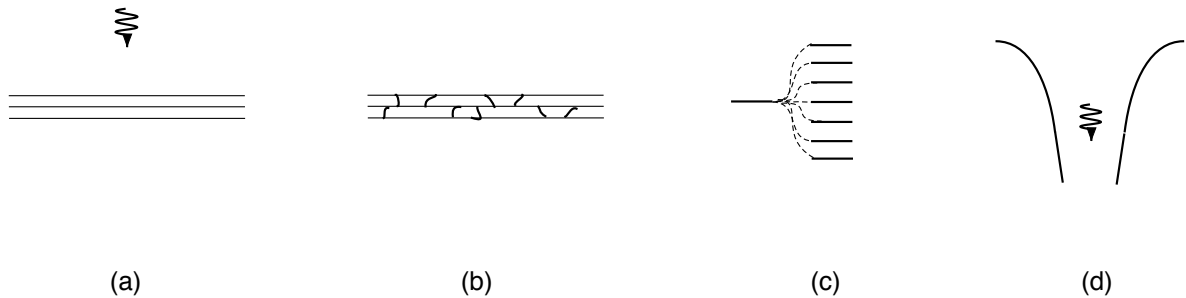}
\caption{\label{f10} (a) A graviton is incident on a stack of D-branes. (b) The graviton is absorbed onto the branes, with its energy getting converted to open strings stretching between the branes. (c) For the D1D5 system this transition can be modeled as absorption from a single level (the graviton) to a closely spaced band of levels (the allowed excitations on the effective string describing the D1D5 system. (d) The corresponding process in the gravity picture: the graviton gets sucked into the `throat' of the metric produced by the branes.}
\end{center}
\end{figure}

(a) Suppose we have a stack of D-branes, and we have a graviton falling towards this stack (fig.\ref{f10}(a)). The graviton hits the branes and gets converted to a collection of open strings (fig.\ref{f10}(b)). One can compute the cross section $\sigma_{branes}$ for this absorption onto the branes.  In particular,  for the bound state of D1 and D5 branes,  $\sigma_{branes}$ can be computed in the following way \cite{interactions}.  The bound state  can be modeled by a long `effective string' \cite{maldasuss}. The energy levels on this string form a  closely spaced band (fig.\ref{f10}(c)). An incident graviton is absorbed onto this string, with the cross section $\sigma_{branes}$  being given by the `fermi golden rule', which describes absorption into a band of levels. 

 On the other hand, we can write down the metric produced by these branes, and compute the cross section $\sigma_{gravity}$ for the graviton to be sucked into the `throat' of this metric (fig.\ref{f10}(c)). One finds that  $\sigma_{branes}=\sigma_{gravity}$ \cite{callanmalda,comparing,interactions}. 

\b

(b) The insight of AdS/CFT \cite{maldacena,gkp,witten} is now the following: the further evolution of the excitation on the effective string can be mapped onto the further progression of the graviton down the throat of the gravity solution. (We are of course being schematic here; the effective string dynamics should really be thought of as a strongly coupled CFT, and the throat region is its gravity dual, an asymptotically AdS space.) 

If the AdS region has a simple geometry, like global AdS, then the AdS/CFT map is straightforward to understand: the excitations in the CFT are dual to excitations around this AdS background.  But what happens if there is a horizon at some depth in the throat? The time $t$  in the CFT  is the analogue of Schwarzschild time, so it goes to infinity as the graviton reaches the horizon. In that case, is the interior of the horizon captured at all in the CFT? We can ask this puzzle in a different way: if the gravity theory does have the region interior to the horizon, then has AdS/CFT failed when black holes form?

In recent years, many different answers have been proposed to this question. In particular, in \cite{cool} it was argued that smooth horizons emerge when the gravity solution is maximally entangled with a second system. We will argue for a different resolution of the puzzle, given by the fuzzball complementarity conjecture \cite{fuzzcomp}: spacetime ends at a fuzzball surface just before the horizon would be reached, and an effective interior region arises only as a description of $E\gg T$ impacts onto this fuzzball surface.

\b

(c) Let us now describe a model that will illustrate this idea of fuzzball complementarity.  As we have seen above in steps (a), (b), the model of fig.\ref{f10}(c) describes the  absorption of a graviton into the throat of the geometry, and its progression upto the vicinity of the horizon.  In the fuzzball picture the spacetime ends just outside the horizon, so there is no `black hole interior' to fall into. But suppose the graviton had a high energy $E\gg T$. When this graviton reaches near the fuzzball surface, the fuzzball state changes, and this change evolves in a certain way.   The conjecture of fuzzball complementarity says that this evolution can be mapped (approximately) onto the radial infall of a quantum in the  {\it vacuum} geometry of  the traditional picture of a black hole. 

\begin{figure}[h]
\begin{center}
 \includegraphics[scale=.82] {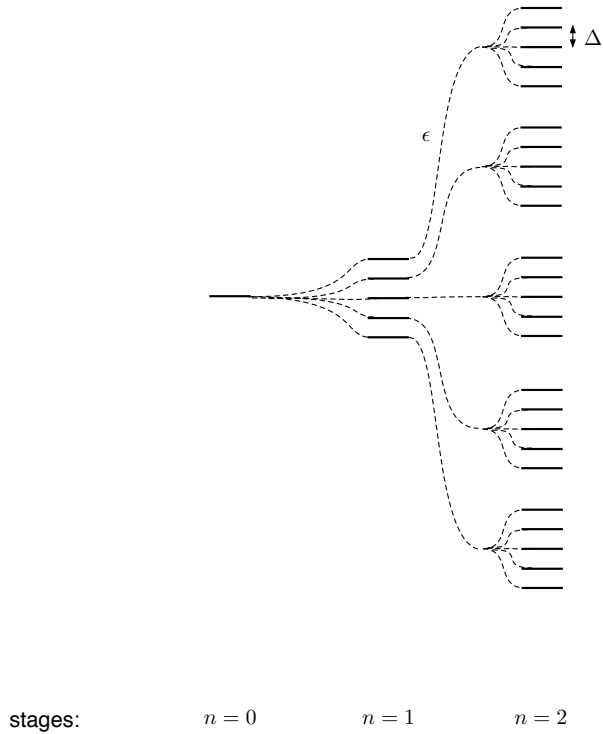}
\caption{\label{ftwo} The energy levels of the CFT are arranged in `stages' $n=0,1,2,\dots$. A level at stage $n$ can transition to a band of levels at stage $n+1$. The  amplitude  for this transition is $\epsilon$  per unit time, and the level spacing is $\Delta$. The Hamiltonian coupling can cause transitions $n\r n+1$ as well as $n+1\r n$, but the overall amplitude drifts towards larger $n$ by the process of `fermi-golden-rule absorption' which occurs whenever the a given energy level can transition into a closely spaced band of levels.    }
\end{center}
\end{figure}

To model this evolution, we assume that  the fuzzball states are arranged in  stages $n$ of increasing  `complexity', as indicated in fig.\ref{ftwo}. The states at stages $n=0$ and $n=1$ are the same as the states depicted in fig.\ref{f10}(c); thus the single level at stage $n=0$ is the graviton incident from asymptotically  flat space, and the band at stage $n=1$ gives the levels used to model absorption into the throat and progression down the throat upto the vicinity of the horizon. The states in stages $n=2, 3, \dots$ describe different fuzzball states that live near the horizon. Thus these states are accessed only when the graviton falls down the throat to a location near the fuzzball surface. 

\b

(d) In this model, a state at any stage $n$ gets `pulled' into a band of states at stage $n+1$ by a `fermi golden rule' transition similar to the one that  gave the transition from stage $n=0$ to stage $n=1$.  We will find that if we start with a nonzero amplitude in stage $n=0$ (and no amplitude in other stages), then  evolution over time pulls the amplitude towards larger and larger values of $n$. The evolution towards increasing $n$, for $n\ge 2$, is mapped to the notion of falling deeper  into the interior of the horizon.

\b

\begin{figure}[h]
\begin{center}
 \includegraphics[scale=.82] {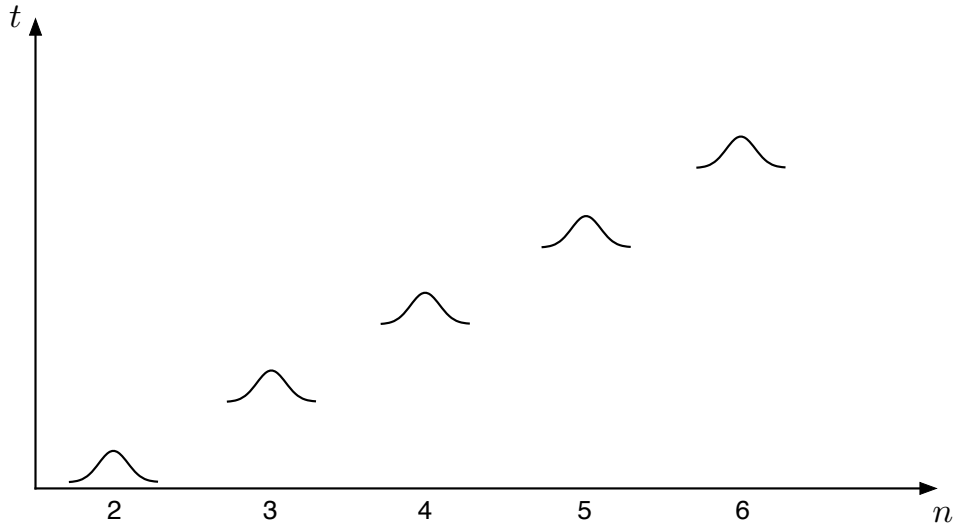}
\caption{\label{fevolution} The horizontal axis label the stage $n$ in our filtration of fuzzball states. The vertical axis is CFT time $t$. The wavefunction peak evolves towards larger $n$ as given by (\ref{main}). This evolution mimics movement through an emergent direction along $n$, which we identify with infall inside the horizon towards the center of the hole.}
\end{center}
\end{figure}

(e) To illustrate the nature of the evolution we expect, we make a toy model as follows. We choose simple values for the inter-level spacings in the bands (all spacings equal to $\Delta$), and simple values for the transition amplitudes (all transition amplitudes equal to $\epsilon$). For this setup, we solve for the evolution of amplitudes in closed form. At time $t$, we find that the amplitudes peaks at the stage $n(t)$ given by
\be
n(t)={2\pi |\epsilon|^2 t\over \Delta }
\label{main}
\ee
so we see that the amplitude indeed gets pulled towards larger $n$. We depict this evolution in fig.\ref{fevolution}. 

Note that the probability to transition from any state $\psi_1$ to a state $\psi_2$  is equal to the probability to transition from $\psi_2$ back to $\psi_1$. The evolution towards larger $n$  arises  from the fact that there is a larger number of states at larger $n$. Thus we need a high density of states to get the picture conjectured here. For this reason we can say that this effective emergence of the black hole interior is a result of the fact that the black hole possesses  a large entropy $S_{bek}$.

\b

\b

(f) Finally we relate the nature of the above model to the physics of fuzzballs. If we  look only at the physical 3+1 dimensional spacetime, then we cannot really understand how any kind of complementarity can emerge. Instead, we note that  the evolution should be studied in {\it superspace} -- the space of all fuzzball states $|F_i\rangle$ \cite{tunnel}
\be
\Psi=\sum_j C_j |F_j\rangle ~\r ~ \sum_j C_j e^{-iE_j t}|F_j\rangle
\ee
 The states on this superspace can be arranged in order of increasing complexity, and evolution in the direction of this complexity is what we are regarding (in a dual description) as infall into the traditional hole. In \cite{as} this notion of states with different complexity was explained in the dual CFT. This CFT is a `symmetric orbifold' CFT in 1+1 dimensions. An infalling graviton starts in the `symmetric' sector of states, and remains in this sector until (in the gravity dual) it approaches a horizon. But on reaching near the horizon, the amplitude in the CFT state starts flowing towards `antisymmetric' sectors of the CFT, and this corresponds to the increase in complexity we are modelling.

\b

(g) Let us now put our result in the context of the different pictures that have been proposed for the quantum physics of black holes:

\b

(i) {\bf The fuzzball paradigm:} In this paradigm, the black hole microstates do not have the traditional vacuum horizon; the spacetime ends in a quantum mess just  outside the place where the horizon would have formed, and the fuzzball radiates from its surface like a star. The conjecture of fuzzball complementarity then seeks to recover the traditional black hole interior as an effective dynamics for the impact of $E\gg T$ quanta onto the fuzzball. The model of the present paper illustrates this conjecture. This paradigm uses string theory as the theory of gravity, since features of this theory are involved in bypassing the no hair theorem and arriving at the microstate construction \cite{gibbonswarner}.

\b

(ii) {\bf The firewall paradigm:} In this approach one does not seek to give an actual mechanism  to alter the vacuum state at the horizon. Instead one makes a generic argument, valid for all theories of gravity, along the following lines. {\it If} there is a modification of the dynamics of the hole which solves the problem of growing entanglement, then one cannot have {\it any} effective description of the black hole interior that mimics the vacuum. One does not make any approximation $E\gg T$; on the contrary, the firewall argument is based on looking at the $E\sim T$ modes that will evolve to Hawking radiation. The argument  uses the bit model and strong subadditivity tools used in \cite{cern}, but with the additional assumption that any effects that deviate from the semiclassically known physics are confined to the interior of the stretched horizon. 

The difficulties with the assumptions used in the firewall argument were discussed in \cite{mt1,mt2}. The model of the present paper shows how the firewall argument can be  bypassed by fuzzball complementarity: once we require that complementarity emerge only in the limit $E\gg T$, then one can have a complementary description of the infall and thus no firewall. 

\b

(iii) {\bf The wormhole paradigm:} In this paradigm one keeps the horizon to be a vacuum region. One then bypasses the Hawking problem by making a new postulate: the Hilbert space at infinity is not independent of the Hilbert space inside the hole. In the approach of Maldacena and Susskind \cite{cool} this nonlocal identification identification of Hilbert spaces 
is achieved by a wormhole connecting the radiated quanta back to the black hole interior. In \cite{pr} no specific model for the identification is proposed, but the consequences of the nonlocal identification for complementarity are analyzed. The difficulties with this paradigm were discussed in \cite{entangled}.

In the context of the wormhole paradigm, the model of the present paper shows that we do not need such a nonlocal identification of Hilbert spaces to get an effective complementarity, provided we require this complementarity only in the limit  $E\gg T$.

\section{Modelling absorption by D-branes}

We will begin by recalling a simple model of absorption by D-branes, where an incident graviton is absorbed  into a band of closely spaced levels on the branes.  Next we recall the idea of symmetric and anti-symmetric sectors in the CFT description of the branes. We explain why the infalling graviton will start in the symmetric sector, but evolve towards the antisymmetric sector.  Finally, we put these two notions together to make a model of absorption where we have a sequence of bands arranged in stages $n=0,1,2, \dots$, and evolution takes a state in the stages $n=0,1$ (corresponding to the symmetric sector) towards larger values of $n$, which correspond to the antisymmetric sector.

\subsection{Absorption into a single band}

Consider the D1D5 bound state system introduced in \cite{sv,callanmalda}. We compactify 10-d type IIB string theory as $M_{9,1}\r M_{4,1}\times S^1\times T^4$. We wrap $n_1$ D1 branes on $S^1$, and $n_5$ D5 branes on $T^4\times S^1$. The bound state of these branes can be modeled as an `effective string' \cite{maldasuss} which has winding number $N=n_1n_5$ around the $S^1$. 

We wish to start with a graviton $h_{ij}$, where $i,j=1, \dots 4$ are directions in the $T^4$. We regularize the noncompact space in $M_{4,1}$ by letting space be  a 4-dimensional box with a large volume $V_{nc}$. Quantizing the graviton field, we write the initial state of this graviton as 
\be
|\psi_g\rangle = {1\over \sqrt{V_{nc}}}{1\over \sqrt{2\omega}} e^{i\vec k \cdot \vec x -i \omega t}\hat A^\dagger_{\vec k, ij}|0\rangle, ~~~~~~~~~~~~~~~~\omega=|\vec k|
\label{gravitonlevel}
\ee
where $\hat A^\dagger_{\vec k, ij}$ is the creation operator for the graviton $h_{ij}$ with wavenumber $\vec k$. 

The graviton will be absorbed onto the effective string, so that its energy will be converted to left and right moving vibrations of  this string. These vibrations can be in any of the $T^4$ directions $i=1, \dots 4$. An excited state of the string has the form
\be
|\psi_{s}\rangle = \left ( {1\over \sqrt{L_{eff}}}{1\over \sqrt{2\omega'}}e^{i\omega'(y- t)}\hat a^{\dagger,L}_{\omega',i}|0_L\rangle\right ) \left ( {1\over \sqrt{L_{eff}}}{1\over \sqrt{2\omega'}}e^{i\omega'(-y-t)}\hat a^{\dagger,R}_{\omega',j}|0_R\rangle\right )
\label{stringlevel}
\ee
Here $y$ is a coordinate along the effective string, which wraps along the $S^1$ direction of spacetime. $L_{eff}$ is the length of this effective string. Assuming that this string is `multiwound' around $S^1$ to make a single string, we find that $L_{eff}=n_1n_5L$, where $L$ is the length of the $S^1$.  The left and right movers have equal momenta $|p|=\omega'$ , since the incident graviton had  no momentum along $y$. The operator $a^{\dagger,L}_{\omega',i}$ creates a left moving vibration on the string with energy $\omega'$ and polarization $i$; similarly for $a^{\dagger,R}_{\omega',j}$.

The energy levels on the string are
\be
E_n=E_n^L+E_n^R={2\pi n\over L_{eff}}+{2\pi n\over L_{eff}}={4\pi n\over L_{eff}}
\ee
The crucial point is that the level spacing 
\be
\Delta = {4\pi\over L_{eff}}={4\pi \over n_1n_5 L}
\ee
is very small, since $n_1, n_5$ are assumed to be large. Thus the energy levels on the effective string form an almost continuous band.  In \cite{interactions} it was found that the amplitude per unit time to transition from the state (\ref{gravitonlevel}) to the string state (\ref{stringlevel}) is
\be
R =  {\sqrt{2}\, \kappa\omega'\over \sqrt{2\omega}\sqrt{V_{nc}}}
\ee
where the Newton constant is $G_N=2\kappa^2$. Starting with the state $|\psi_g\rangle$ at time $t=0$, we find that the amplitude in the state $|\psi_s\rangle$ at time $t$ is given by
\be
A(t)=R e^{-iE_n t}\int_0^t dt' e^{i(E_n-\omega)t'}=Re^{-{i\over 2}(E_n+\omega_0)t} \left (  {2 \sin [(E_n-\omega)t/2]\over (E_n-\omega)}\right ) 
\ee
The total probability of absorption at time $t$  into the states on the string is
\be
P(t)=\sum_n |R|^2 \left (  {2 \sin [(E_n-\omega)t/2]\over (E_n-\omega)}\right ) 
\ee
Since the energy levels have a small spacing $\Delta$, we replace the sum by an integral 
\be
\sum_n \r \int {dE\over \Delta}
\ee
The probability of absorption per unit time then becomes
\be
{\cal R}_{string}={P(t)\over t}={2\pi |R|^2\over \Delta}\times 2 = {4\pi |R|^2\over \Delta}
\label{prob}
\ee
where the extra factor of $2$ comes from the fact that $h_{ij}$ can also be absorbed in to the state which has  the $i,j$ indices interchanged in (\ref{stringlevel}).  

One then finds that this rate of absorption agrees exactly with the rate of absorption ${\cal R}_{geometry}$ of the graviton $h_{ij}$ into the geometry produced by the D-branes. (fig.\ref{f10}):
\be
{\cal R}_{string}={\cal R}_{geometry}
\ee
 
\subsection{Phase Analysis of absorption into a band}

\begin{figure}[h]
\begin{center}
 \includegraphics[scale=.60] {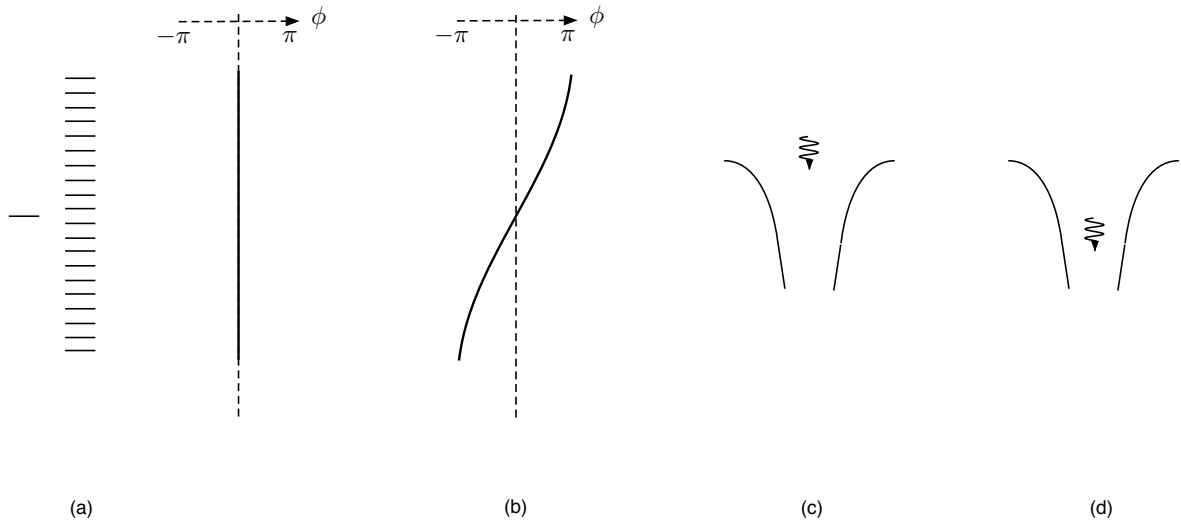}
\caption{\label{fphases2} (a) The single level on the left represents the graviton state $|\psi_g\rangle$. This state can transition to the states on the string  $|\psi_s\rangle$; these states are indicated by the band of levels. The initial amplitude attained on all these states of the band is the same; we have called it $\phi=0$. (b) After some time $t$, the phases of the states  $|\psi_s\rangle$ evolve to become unequal, since these states have different energies. Even though each state in the band can transition back to  $|\psi_g\rangle$, the different phases cause a cancellation of this transition amplitude, so the energy stays in the band for long times. (c) The gravity dual for situation (a); the quantum has just been absorbed into the throat. (d) The gravity dual for situation (b); the quantum has progressed down the AdS throat.}
\end{center}
\end{figure}

Let us recall  the physics of such `fermi golden rule' absorption into a band. To see the significance of having a band, first 
consider the situation where there is only one state $|\psi_g\rangle$ and only one state $|\psi_s\rangle$ on the string. Then the amplitude would oscillate periodically between these two states, instead of moving monotonically from $|\psi_g\rangle$ towards $|\psi_s\rangle$. Now consider the case where we have the band of levels on the string. The absorption will be peaked around $E_n =2\omega'\approx\omega$, so we may approximate the transition amplitude $R$ by a constant for all levels. Thus the initial state $|\psi_g\rangle$ will transition to all the energy levels $|\psi_s\rangle$ with the same phase; we depict this in fig.\ref{fphases2}(a) by indicating the same phase $\phi=0$ for all levels.   

At later times, the  evolution of these energy levels changes these phases so that they become unequal, as depicted in fig.\ref{fphases2}(b).  The levels with $E_n>\omega$ move towards phases $\phi>0$, and the levels $E_n<0$ towards phases $\phi<0$. 

Thus even though the states $|\psi_s\rangle$ on the strings can transfer amplitude back to the state $|\psi_g\rangle$ with the  amplitude per unit time  $R^*$, the amplitude in the states $|\psi_s\rangle$ have different phases, and so these contributions cancel. This is the reason that the amplitude flows from the state $|\psi_g\rangle$ to the string, but does not  return back to the state $|\psi_g\rangle$ for a long time. If the levels are evenly spaced, then the phases in the states $|\psi_s\rangle$ will become equal again after a time $t\approx {2\pi\over \Delta}$, but since $\Delta $ is small, this is a very long time. More generally, the spacing between levels will be uneven, with irrational ratios for the spacings between different levels. In this case the phases of all the states $|\psi_s\rangle$ will never agree after $t=0$, and an approximate return to the state $|\psi_g\rangle$ can take even longer. 

\subsection{The AdS/CFT duality map}\label{adscft}

Once the graviton is absorbed onto the string. the state on the string continues to evolve.  The very nice insight of AdS/CFT duality \cite{maldacena,gkp,witten} is that this phase of the evolution is dual to the further progression of the graviton down the `throat' of the geometry -- a region where the metric is locally AdS. This is depicted in figs.\ref{fphases2}(c),(d). Of course the field theory on the strings is not as simple as the free theory we have used; it is a strongly coupled CFT. We will need to look at the interactions in this CFT later, but for now  the free model will serve to illustrate the general idea we are trying to describe.

\begin{figure}[h]
\begin{center}
 \includegraphics[scale=.60] {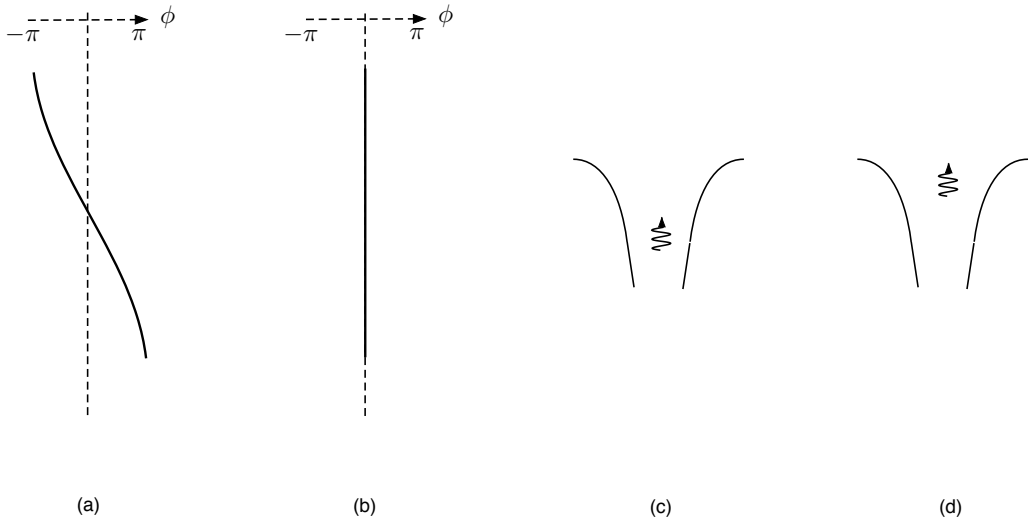}
\caption{\label{fphases3} Showing that the motion of fig.\ref{fphases2} is reversible: (a) We start with phases in the CFT that are positive for lower energies, and negative for higher energies. (b) These phases evolve to the situation where the phases are all zero. (c) The gravity dual of (a) is a quantum in the throat moving upwards. (d) The gravity dual of (b); the quantum has reached the top of the throat.}
\end{center}
\end{figure}

It is very important to note that motion in the AdS throat is `reversible'. In fig.\ref{fphases3} we depict a graviton in the AdS throat, going {\it up }. In the CFT this corresponds to starting with the phases as shown in fig.\ref{fphases3}(a). The phases of higher energy states evolve towards more positive values, and so after a time the phases become all equal, as in fig.\ref{fphases3}(b). This evolution of phases in the CFT state corresponds to the motion of the graviton from the location in fig.\ref{fphases3}(c) to the top of the throat as depicted in fig.\ref{fphases3}(d). Further evolution of the phases brings us back to the phases of fig.\ref{fphases2}(b); this corresponds to the graviton reflecting off the AdS boundary and falling back down the throat. 

There is a small probability that the graviton exits the throat -- this is equal to  the probability (\ref{prob}) of being absorbed into the throat from the asymptotically flat spacetime. But in the limit of low energies for the graviton, this probability is small, and we can take the decoupling limit where the low energy excitations of the CFT on the branes is exactly dual to the gravity theory in the AdS region \cite{maldacena}.

\begin{figure}[h]
\begin{center}
 \includegraphics[scale=.82] {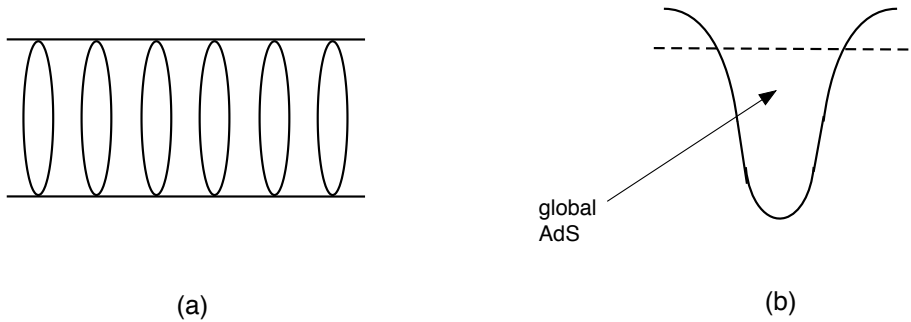}
\caption{\label{globalads} (a) In the CFT state the effective string is broken up into $n_1n_5$ separately wound circles. (b) The geometry below the dotted line (which excludes the flat space and neck regions) is global $AdS_3\times S^1\times T^4$; this is the gravity dual of the CFT state (a). }
\end{center}
\end{figure}

In the above discussion we have looked at only the `throat' region of the AdS geometry. What happens if we keep going down this throat? In \cite{lm4} it was shown that this throat is always `capped', with the shape of the cap depending on the choice of CFT state. As an example let us take the simplest state, where the throat is capped off in such a way that the cap+throat region gives global AdS \cite{bal,mm}. The dual CFT state has the following structure. Recall that the `effective string' made from the D1 and D5 branes had a total winding number $N=n_1n_5$.  But we need not take this effective string to be in the form of one `multiwound' string. Instead, we can let the effective string form $N$ separate `singly wound loops'. (Each separate loop is called a `component string'.)  This CFT configuration is   depicted in fig.\ref{globalads}; and it is dual, after a change of coordinates, to the spacetime $AdS_3\times S^3\times T^4$.

\begin{figure}[h]
\begin{center}
 \includegraphics[scale=.82] {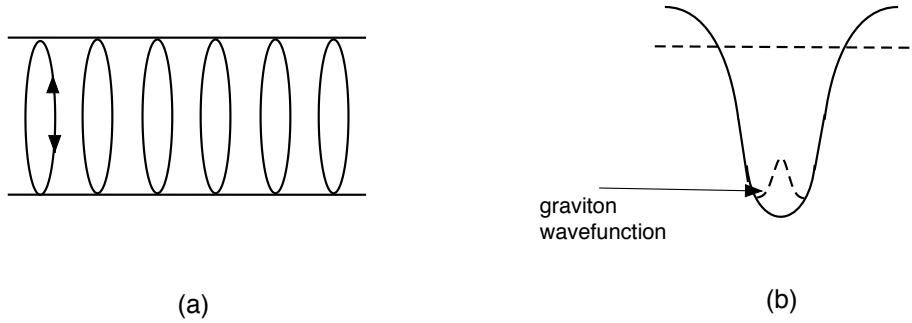}
\caption{\label{globalads2} (a) In the CFT state one of the component strings has been excited with a left moving and a right moving vibration. (b) In the dual geometry we have a graviton wavefunction localized in the cap region. }
\end{center}
\end{figure}

The excitations of the CFT state  are again of the form (\ref{stringlevel}), but now $L_{eff}=L$. The energy levels are then
\be
E^{CFT}_n={4\pi\over L}, ~~~n=1, 2, \dots
\ee
In the dual AdS space, one can solve the wavefunction equation for the graviton, getting energy levels for wavefunctions localized near the origin of AdS. One finds exactly the same energy levels $E^{gravity}_n$
\be
E^{CFT}_n=E^{gravity}_n
\ee
 thus giving an illustration of AdS/CFT duality. The excited CFT state and its gravity dual are depicted in fig.\ref{globalads2}.
 
But thus does not solve Hawking's puzzle. We can add enough energy in the AdS space to make a black hole \cite{wittenthermal}. Entangled pairs will be created at the horizon of this hole, and we will find the information problem once again. 

In \cite{eternal} it was suggested that small corrections to Hawking's leading order computation may remove the problematic entanglement. But in \cite{cern} it was shown, using the powerful tool of quantum strong subadditivity, that this is not the case; one needs corrections of order {\it unity} for the evolution of low energy modes at the horizon. 

A further difficulty stems from the fact that with the traditional geometry of the hole, the Schwarzschild time coordinate covers only the exterior of the hole. The time in the CFT appears to be similar to this Schwarzschild time, so one might wonder if the interior of the hole is not captured by the dual CFT.

We will now explain how these issues are handled in the fuzzball paradigm.

\subsection{Symmetric and antisymmetric sectors in the CFT}\label{sa}

\begin{figure}[h]
\begin{center}
 \includegraphics[scale=.82] {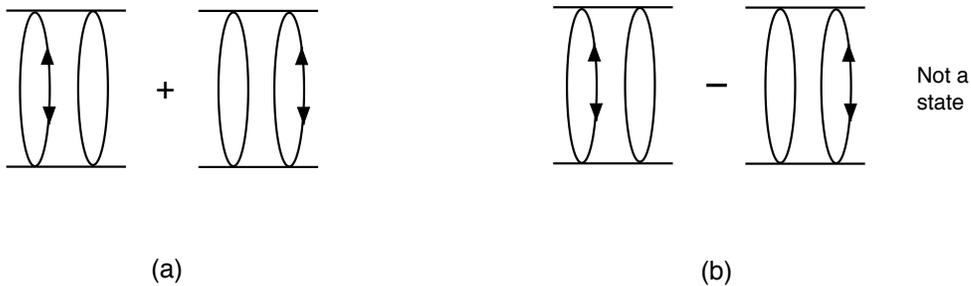}
\caption{\label{fthree} (a) In a symmetric orbifold theory, we must symmetrize the excitation over component strings that are identical. (b) The linear combination  which is antisymmetric  between two identical component strings; such a  state does not exist in the orbifold theory.}
\end{center}
\end{figure}

Consider the CFT state fig.\ref{globalads}(a) dual to AdS space. Let us draw just two component strings out of all the $n_1n_5$ component strings in this state.  We can take an excitation where one of the component strings is excited in the form (\ref{stringlevel}). But because the CFT is a `symmetric orbifold CFT', we have to take the symmetric combination where either of the two strings is excited in this way; this is the state depicted in fig.\ref{fthree}(a). We {\it cannot} take the state in fig.\ref{fthree}(b), since this is not symmetric between the two component strings. 

\begin{figure}[h]
\begin{center}
 \includegraphics[scale=.82] {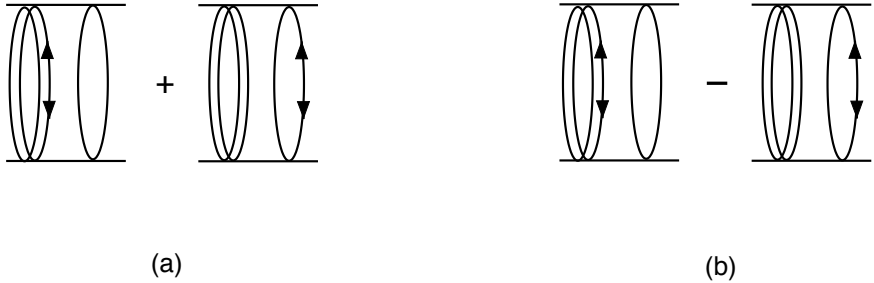}
\caption{\label{ffour}  A case where we have two  component strings with unequal winding. Now both the symmetric (a) and antisymmetric (b) linear combinations are allowed states.}
\end{center}
\end{figure}

But now consider more general states of the D1D5 CFT, where the component strings have different possible winding numbers $N_i$, with
\be
\sum_i N_i=N=n_1n_5
\label{general}
\ee
In fig.\ref{ffour}(a) we depict two component strings out of this collection, one with winding $N_1=1$ and one with winding $N_2=2$. Now we can have the symmetric excitation of fig.\ref{ffour}(a), but we can {\it also} have the antisymmetric excitation of fig.\ref{ffour}(b). This antisymmetric excitation is possible because the component strings are not identical, and so we do not have to symmetrize among them.

The question now is: what are the gravity duals of symmetric and antisymmetric excitations?

Before we add any excitations, we have all the different ground states of the Ramond sector of the D1D5 system. These states are described by different choices of $\{ N_i\}$ in (\ref{general}). The gravity duals of these CFT states were computed in \cite{lm4, twochargeother}. It was found that none of the dual gravity solutions had a horizon or singularity.  Each solution had  the same behavior  of flat space at infinity, a neck and an AdS throat, but a different shape to the `cap'.

  For the simple case where all component strings are the same, as in fig.\ref{globalads2}(a), the geometry is axially symmetric, like the one shown in fig.\ref{globalads}(b). The excitations about such a state are all in the symmetric sector (since all component strings are identical). Their gravity dual excitations are simple as well, like the state shown in  fig.\ref{globalads2}(b).

For the more general case (\ref{general}) of unequal component strings, the gravity dual has a complicated cap structure. Now the cap is not axially symmetric, and can support a large number of different wavefunctions. These wavefunctions are of two types:

\begin{figure}[h]
\begin{center}
 \includegraphics[scale=.82] {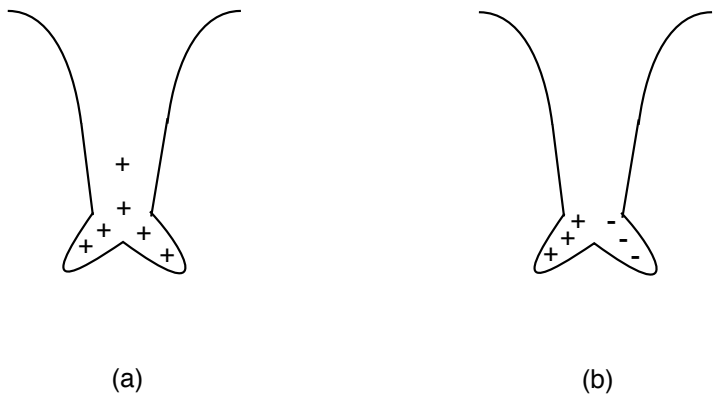}
\caption{\label{diffcaps} (a) A wavefunction in the symmetric sector; the $+$ signs indicate the phase of the wavefunction. It is the same in both branches of the cap, and so there is a nonzero tail up into the AdS region. (b) A wavefunction in the antisymmetric sector; the phase is opposite in the two branches of the cap, so the wavefunction does not extend up into the AdS region.}
\end{center}
\end{figure}

\b

(i) The symmetric CFT state of fig.\ref{ffour}(a) corresponds to graviton wavefunctions like the one in fig.\ref{diffcaps}(a); the wavefunction is symmetric in the two branches of the cap, and has a tail that extends towards to boundary of the AdS region.

\b

(ii) The antisymmetric CFT state of fig.\ref{ffour}(b) is dual to a wavefunction like the one in fig.\ref{diffcaps}(b); the wavefunction has opposite signs in the two branches of the cap, and there is virtually no `tail' leaking towards the boundary of AdS. The wavefunction from each branch of the cap does tend  to leak into the upper part of the throat, but because the wavefunction has opposite signs in the two branches, this leaked part {\it cancels}. 

\b

Thus the antisymmetric sector states should be thought of as states that are confined to the vicinity of the cap region of the geometry, while the symmetric sector states can extend all the way to the boundary of AdS.

Because antisymmetric modes are localized in the cap, they do not connect directly to the asymptotically flat spacetime outside the geometry. Thus they do not directly contribute to the radiation from the D1D5 geometry; the energy must first leak to the symmetric sector modes, and these modes will extend to the neck and leak out to the flat space region. This fact was noted in the CFT description in \cite{falling}, where emission from a general CFT state (\ref{general}) was considered.  Each of the excited component strings can emit into the same graviton mode $h_{ij}$, but because the excitations on these different strings have different phases  (for a state not in the symmetric sector), these emissions `phase cancel', and the emission of the graviton is highly suppressed.

Let us now put together the above picture of symmetric and antisymmetric modes into the evolution problem that we wish to study:

\b

(A) The general extremal state of the D1D5 system (\ref{general}) has a dual geometry with the structure in fig.\ref{diffcaps}. We have flat space at infinity, then a locally AdS throat, and then a cap whose detailed structure depends on the choice of state from the set (\ref{general}). 

\b

(B) Now consider a graviton that comes in from infinity and is absorbed into the throat of the geometry. In the CFT, the initial excitation this creates is in the {\it symmetric} sector. The reason is the coupling $\sim h_{ij}\p X^i \bar p X^j$  which creates the excitation: this coupling focuses only around  a single point of any component of the effective string, and so it knows nothing about which winding  state (\ref{general}) we have. Thus all strands of the effective string are excited with the same amplitude, and this leads to an excitation in the symmetric sector.

\b

(C) The graviton now progresses down the throat of the geometry. In the dual CFT, the excitation begins to spread over a larger region on the effective string. Note that the interactions in the CFT are symmetric between all strands, and the initial state is in the symmetric sector, as noted in (B). Thus the state in the CFT remains in the symmetric sector as it spreads, for this part of the evolution. 

\b

(D) At some point the graviton reaches the `cap' region. To start with the gravity state is in the symmetric sector, so it spreads to symmetric states like the one in fig.\ref{ffour}(a). But once the graviton is in the cap region, there is a nonzero amplitude for its wavefunction to also spread to the  the antisymmetric sector. In the dual CFT the excitation has now spread to wavelengths where it explores the entire length of the effective string, and can distinguish between the states (\ref{general}). There is a slow but steady spread to the antisymmetric sector of CFT states like the one in fig.\ref{ffour}(b), mirroring the corresponding spread to antisymmetric states in the gravity description. 

\b

(E) In \cite{lm5} it was noted that states stay trapped for long times in the cap, so we reproduce in a sense the behavior of an effective horizon: it is easy for the graviton to fall into the cap region, but difficult for it to leak out. Nevertheless the entire evolution is unitary, with no real horizon or singularity. Our goal is to make a model for the progressive evolution of states from the symmetric to the antisymmetric sector; we will regard this progressive evolution as a dual description of infall into the interior of a traditional hole. Note that this evolution happens after the graviton reaches near the boundary of the `fuzzball', so we are examining an evolution  in the space of fuzzball excitations. 

\subsection{Modeling the evolution of fuzzball states}

In the above discussion we were looking at low energy excitations of 2-charge extremal D1D5 states; these extremal states are described by different windings (\ref{general}) of the effective string, and we have looked at the infall of a single graviton towards such states. We will actually be interested in states with large excitations, which make near-extremal black holes. The dynamics of such states will be discussed in later sections, but the model that we wish to propose for fuzzball complementarity can already be explained with the simple notions we have discussed so far. Let us therefore discuss the model first, and return later to the discussion of how it describes fuzzball complementarity.

We consider our model in the following steps, which roughly parallel the steps (A)-(E) in section \ref{sa}:

\b

(A') We start with the 2-charge extremal geometry, and a graviton outside the throat. The state of the graviton is  modeled by a single energy level; we will call this level `stage $n=0$'. At time $t=0$ we start with amplitude $f^{(0)}=1$ in the level at stage $0$. This stage $n=0$ is depicted in fig.\ref{ftwo}. 

\b

(B') Now consider D1D5 system in one of the states (\ref{general}). The effective string has a closely spaced band of levels. We depict this band as the levels at stage $n=1$ (fig.\ref{ftwo}). We introduce a small coupling $\epsilon$ between the level at stage $n=0$ and any level of stage $n=1$. This coupling corresponds to the interaction  that couples the graviton to excitations of the effective string. At time $t=0$ there is no amplitude in any of the levels of stage $n=1$;  we write this as $f^{(1)}_j=0$. But as $t$ increases, the amplitude will pass from the level at stage $n=0$ to the levels in stage $n=1$; this is just the absorption onto the effective string that we discussed above, and corresponds to the graviton being absorbed into the throat of the geometry produced by the branes (fig.\ref{f10}(d)).

\b

(C') The amplitudes $f^{(1)}_j$ in stage $n=1$ evolve as $\sim e^{-iE_j t}$; This gives the phases depicted in fig.\ref{fphases2}(b), and corresponds to evolution in the CFT. In the gravity picture, the graviton progresses down the throat of the dual geometry (fig.\ref{fphases2}(d)).

\b

(D') Upto  this point the model was similar to the one used in \cite{interactions} for absorption by D-branes. We now add in the new feature that each level at stage $n=1$ can itself transition into a new band of levels (fog.\ref{ftwo}). This corresponds to the fact that a state in the symmetric sector (stage $n=1$) transition to states that are  in the antisymmetric sector (stage $n\ge 2$). For our simple model we take the same spacing $\Delta$ for all levels in all stages, and also the same coupling $\epsilon$ between a level at any stage and the next. On the gravity side, this transition from amplitudes $f^{(1)}_j$ to amplitudes $f^{(2)}_{jk}$ at stage $n=2$ corresponds to the change from symmetric wavefunctions (which extend up the throat of the geometry) to antisymmetric wavefunctions that live only near the horizon region (fig.\ref{diffcaps}(b)).  

\b

(E') Each state at stage $n=2$ has a similarly allowed transition into a band of levels at stage $n=3$, and so on. This corresponds to the CFT states moving `deeper into the antisymmetric sector'. Thus we visualize the states as being arrange in an order of `complexity': the simplest ones are the symmetric sector states, which are naturally accessed by a graviton falling down the throat (stage $n=1$ states). Then we have states that can be accessed from these $n=1$ states; these are termed stage $n=2$ states. Then we have states that can be accessed only from the stage $n=2$ states; these are stage $n=3$ states, and so on. This pattern of states is normal for any sufficiently complicated system (though an effective new coordinate like that depicted in fig.\ref{fevolution} does not have to emerge in every case). For example, suppose we take a gas in a box, and start with all the atoms having a the same velocity $v$ and velocity only in the $x$ direction. After one collision we will access a larger number of states, after two collisions we will access yet more states, and so on till we reach a generic thermalized state. In our problem, the evolution to higher stages $n$ corresponds to progressive thermalization in the CFT; similarly, in the gravity description, we are moving towards a generic state $\sum_i C_i |F_i\rangle$ in the space of fuzzball states. This progression to higher stages is depicted in fig.\ref{fevolution}, where it mimics infall through an emergent radial direction. 

\b

\subsection{The physical picture  emerging from the model}

Let us now comment on the physics that we wish to describe through such a model:

\b

(1) {\it Irreversibility:} As we had noted in section \ref{adscft}, the physics of the AdS {\it throat} region is reversible: we have states that progress down the throat, and states the progress up the throat (fig.\ref{fphases2} and fig.\ref{fphases3}). But the physics we see now is the physics of the {\it horizon}, which should appear `irreversible' in some leading order analysis. More precisely, a high energy quantum ($E\gg T$)  should be able to fall into the horizon, but not emerge the same way; its energy and information should leak out only slowly through low energy $E\sim T$ quanta. 

From fig.\ref{ftwo} we can see how this irreversibility arises in our model. It is easy for amplitudes to move towards higher stages $n$, since there are more states at higher $n$. For the same reason, there is a very low probability to head in the opposite direction; i.e., towards lower $n$. Thus if we regard the direction of increasing $n$ as the direction of infall through the horizon, then we see that it is easy to fall into the horizon but difficult to get out.

\b

(2) {\it Emergence of spacetime:} Of course any sufficiently complicated system would have a behavior where it is possible for high energy quanta  to come in, but hard for high energy quanta to leave. This is after all the second law of thermodynamics, which says that heat comes into a system at a high temperature,  leads to an irreversible increase in entropy, and leaves at a lower temperature.  The extra point here is that we are identifying the progression through the stages $n$ as infall along an emergent coordinate direction $r$, which ranges from the horizon to the singularity. Thus the question arises: when can we identify the evolution in the space of states with infall along an emergent spatial direction? Consider the following situations:

\b

 \begin{figure}[h]
\begin{center}
 \includegraphics[scale=.82] {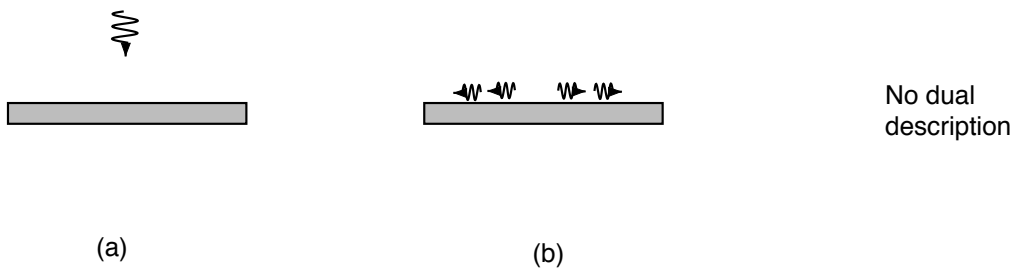}
\caption{\label{concrete} (a) A photon is incident on a slab of metal (b) The photon gets abosrbed into the slab, with its energy converted to waves in the metal. There is no dual description where the incident photon passes smoothly through the place where the metal slab exists.}
\end{center}
\end{figure}

(i) In fig.\ref{concrete}(a) we depict a photon impinging on a metal slab. The photon is absorbed, with its energy going towards exciting the motion of electrons in the slab (fig.\ref{concrete}(b)). The energy on the metal slab spreads away from the point of impact; this is the future evolution of the system. But there is no `dual' description of this evolution where the photon continues unimpeded through the point where it hits the slab.

\b

(ii) Now recall fig.\ref{f10}(a) where we had a graviton impinging on a stack of D3-branes. As depicted in fig.\ref{f10}(b), The graviton is absorbed, with its energy getting converted to gluons -- open strings extending between the D3-branes. Thus far, it looks similar to case (i); the energy on the D3-branes spreads out from the point of impact. 

\b

(iii) In fig.\ref{f10}(d) we see that this process has a dual description; in this latter description the graviton  continues unimpeded through the place where it had hit the branes. The dynamics of the branes has been replaced by the dynamics of infall into the AdS region. So unlike case (i), the infalling quantum in (ii) does not get destroyed. Thus is of course the lesson of AdS/CFT duality. But what is the difference between cases (i) and (ii)?

\b

In \cite{fuzzcomp} this difference was explained as follows. First consider case  (i), the case of the metal slab. The incident photon is expressed as some linear combination of eigenstates $|\psi\rangle=\sum_i C_i |E_i\rangle$. After absorption in the slab, let its  wavefunction changes  to $|\psi'\rangle=\sum_i C'_i |E'_i\rangle$, where $|E'_i\rangle$ are eigenstates for the metal slab:
\be
\sum_i C_i |E_i\rangle\r \sum_i C'_i |E'_i\rangle
\label{change}
\ee
 We have in general
\be
C_i\ne C'_i, ~~~~E_i\ne E'_i
\ee
since the locally accessible energy levels $E'_i$ in the metal slab are not the same as the levels $E_i$ in the incident photon. Now consider case (ii), the case of a graviton incident on  large number $N$ of D-branes. We again have the change (\ref{change}) when the graviton breaks up into open strings when it hits the branes. But the level density is so high (because $N\gg 1$) that we get an almost faithful map from $|\psi\rangle\r|\psi'\rangle$:
\be
C_i= C'_i+o({1\over N}), ~~~~E_i=E'_i+o({1\over N})
\label{ce}
\ee
Thus we get a {\it duality}, where the space of states $|\psi\rangle$ allowed for the incident graviton map almost faithfully into the space of states allowed on the D-branes. In the decoupling limit (low energies for the graviton) this map becomes exact if we consider the physics in the AdS region; this is AdS/CFT duality. 

Returning to our problem, there are two relevant aspects:

\b

(a) In getting (\ref{ce}), it was important that the level density on the branes be high, so that every energy level $E_i$ for the graviton find a close match $E'_i$ on the branes. For black holes, the level density is high because the Bekenstein entropy is very large; much larger than the entropy of normal matter. 

\b

(b) The above condition (a) is necessary for getting a duality, but it is of course not sufficient; the absorption of the graviton onto the branes must be `universal', in the sense that the dynamics of absorption should yield $C_i=C'_i+o({1\over N})$ for all energy levels $i$. For AdS/CFT duality this is a {\it conjecture}, which must then be checked by examining the dynamics of AdS and of the CFT. In our case it is a conjecture as well; one must examine the progression of the wavefunction along our hierarchy of stages, and see that it does map approximately onto infall in the interior of the traditional horizon.

\section{Solving for the evolution}

In this section we perform the computation that leads to the result (\ref{main}). We will first set up notation for the system mentioned above, where states at any given stage $n$ can transition into a band of states at the next stage $n+1$. We will then find the progression of the amplitude along the stages $n$. 

\subsection{The setup}\label{setup}

The energy levels in our system will be arranged in `stages'. The excitation will move from lower stages to higher stages in the process of Hamiltonian evolution, and our goal is to track this progression of the excitation. We denote the stages by an integer $n=0, 1, 2, \dots$. Let us now describe the structure of our system in detail:

\b

(a) We let the lowest stage have just one energy level. Since we can shift all energies by a constant $E_k\r E_k+C$, we choose the energy of this level to be $E=0$. The amplitude in this energy level at time $t$ will be denoted $f^{(0)}(t)$. We will start the evolution at $t=0$, with the amplitude
\be
f^{(0)}(0)=1
\label{qone}
\ee
and no amplitude in any of the  levels in other stages.

\b

(b) The next stage ($n=1$) has a set of energy levels $E_k$, with $k=-N, \dots N$; thus there are $2N+1$ levels at this stage. The separation between the  levels is $\Delta$. We will take $\Delta$ to be small and $N$ large, so that we get an almost continuous band of levels. The central level in this band (i.e. the level $E_k$ for $k=0$) has the same energy $E_0=0$ as the energy level in the previous stage. Thus
\be
E_k= k\, \Delta, ~~~-N\le k \le N
\ee
The amplitude at time $t$ in the level $E_k$ will be denoted $f^{(1)}_k(t)$.  

\b

(c) The Hamiltonian coupling between the level $E=0$ of stage $n=0$ and any level $E_k$ of stage $n=1$ is $\epsilon$.\footnote{ Note that a more general model would allow $\epsilon$ to depend on the index $k$ of energy level $E_k$, but we will restrict to the case where $\epsilon$ is a constant for all $k$. In our situation where the $E_k$ form a closely spaced band, the amplitude in the stage $n=0$ at $E=0$ will be peaked in a narrow range around $E_k=0$, and in this situation it is not a bad approximation to assume that  the $\epsilon_k$ equal some constant $\epsilon$ over this narrow range.} We assume that 
\be
\epsilon\ll 1
\ee
and expand all quantities in powers of $\epsilon$.  Thus we write
\be
f(t)=f^{(0),0}(t)+f^{(0),1}(t)+f^{(0),2}(t)+\dots
\ee
where $f^{(0),j}(t)$ is of order $\epsilon^j$.
From (\ref{qone}) we see that
\be
f^{(0),0}=1
\ee

\b

(d) Similarly, we can write
\be
f^{(1)}_k(t)=f^{(1),0}_k(t)+f^{(1),1}_k(t)+f^{(1),2}_k(t)+\dots
\ee
Since we start the evolution with $f^{(1)}_k(0)=0$, we have $f^{(1),0}_k=0$. Thus the leading order contribution to $f^{(1)}_k(t)$ is $f^{(1),1}_k(t)$. We have 
\be
\dot f^{(1),1}_k(t)=-i E_k f^{(1),1}_k(t)-i\epsilon f^{(0),0}(t)=-i E_k f^{(1),1}_k(t)-i\epsilon 
\ee
We can solve this equation by writing it as
\be
e^{-iE_k t}{d\over dt} \left (e^{i E_k t} f^{(1),1}_k(t)\right ) = -i\epsilon 
\ee
which gives
\be
f^{(1),1}_k(t)=(-i\epsilon)e^{-i E_k t} \int_{t_1=0}^t dt_1 e^{iE_k t_1} =-{\epsilon\over  E_k}(1-e^{-i E_k t})
\label{qtwo}
\ee

\b

(e) Eq.(\ref{qtwo}) is just the computation encountered in the standard treatment of the fermi golden rule, where the amplitude $f^{(0)}$ in the initial level (stage $n=0$) transitions into a band of levels (stage $n=1$). In our problem, on the other hand, the amplitude in the level $E_k$ will in turn transition into yet another band of levels, as indicated in fig.\ref{ftwo}. The levels in this latter band will be levels of stage $n=2$. 

At $O(\epsilon^2)$ we have two kinds of terms. The first term is $f^{(0),2}$, the correction to the  amplitude at stage $n=0$. Let us evaluate this term. We have
\be
\dot f^{(0),2}(t)=-i \epsilon^* \sum_k f^{(1),1}_k(t)
\ee
The solution is 
\bea
f^{(0),2}(t)&=&-i \epsilon^*\sum_k\int_0^t dt_1 f^{(1),1}_k(t_1)=-i\epsilon^*\sum_k \left (-{\epsilon\over E_k}\right )\int_0^t dt_1(1-e^{-i E_k t_1})\nn
&=&i|\epsilon|^2\sum_k {1\over E_k}[t-{1\over -i E_k}(e^{-i E_k t}-1)]
\eea
The density of levels is
\be
\rho={1\over \Delta}
\ee
Thus we have
\be
\sum_k ~\r ~ {1\over \Delta E} \int dE = {1\over \Delta} \int dE
\label{sum}
\ee
The limits of integration are
\be
E=\pm {N\Delta}\equiv \pm A
\ee
We will assume that our limits $\Delta \r 0$, $N\r \infty$ are taken in such a way that
\be
A\r \infty
\ee
Thus we get
\be
f^{(0),2}(t)={i|\epsilon|^2\over \Delta}\int_{E=-\infty}^\infty{1\over E}(t+{1\over  iE}(e^{-i E t}-1))
\label{ff}
\ee
We can evaluate this integral by the method of contours, as follows:

\begin{figure}[h]
\begin{center}
 \includegraphics[scale=.82] {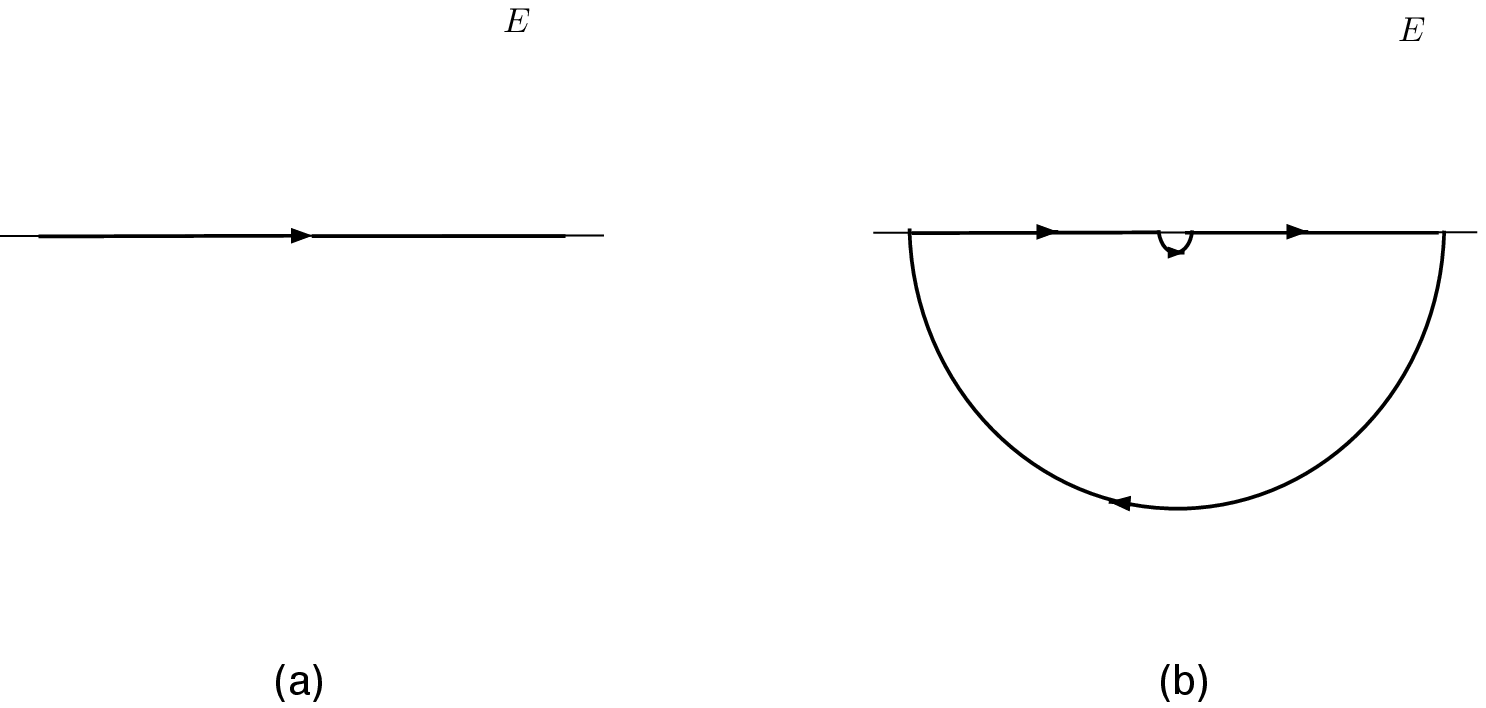}
\caption{\label{fcontour} }
\end{center}
\end{figure}

\b

(i) The integrand has three terms, which together give a function that is nonsingular on the real axis. In particular, even though we have powers of $E$ in the denominator, there is in fact no singularity there in the overall integrand. Thus we slightly deform  the contour to pass below the point $E=0$, as shown in fig.\ref{fcontour}. Since the function is smooth at $E=0$, this makes no difference to the overall integral, but makes it possible to it as a sum  of  three separate terms.

\b

(ii) We close the contour $-\infty <E<\infty$ by adding a semicircle at infinity in the lower half complex plane. This closed contour has no singularities, so the integral around it vanishes. But the semicircle we added does not have a vanishing contribution, and so we must in fact subtract its contribution which we call $C$. We find no contribution from the second and third terms in the integrand, since in each case the integrand falls fast enough at infinity; here we recall we are looking at the evolution for $t>0$. But the first term gives a contribution
\be
C={i|\epsilon|^2\over \Delta} t (-i\pi)=|\epsilon|^2 \pi t 
\ee
Thus
\be
f^{(2),0}(t)=-C=-{|\epsilon|^2 \pi t \over \Delta}
\ee

\b

(f) There is a second term at $O(\epsilon^2)$; this is the term $f^{(2),2}_{jk}(t)$ at stage $n=2$. Recall that this term gives the amplitude in the $k$th level that can be reached from the stage $1$ level $j$. We have
\be
\dot f^{(2),2}_{jk}(t)=-i (E_j+E_k)f^{(2),2}_{jk}(t)-i \epsilon f^{(1),1}_j(t)
\ee
This can be rewritten as
\be
{d\over dt}\left ( f^{(2),2}_{jk}(t)e^{i (E_i+E_j)t}\right )=-i e^{i (E_j+E_k)t}\epsilon f^{(1),1}_j(t)
\ee
We get
\bea 
 f^{(2),2}_{jk}(t)&=&-i\epsilon  e^{-i (E_j+E_k)t}\int_{t_1=0}^t dt_1e^{i (E_j+E_k)t_1}f^{(1),1}_j(t_1)\nn
 &=&-i\epsilon\left ( -{\epsilon\over E_j}\right )  e^{-i (E_j+E_k)t}\int_{t_1=0}^t dt_1e^{i (E_j+E_k)t_1}(1-e^{-i E_j t_1})\nn
 &=&\epsilon^2\left ({1\over E_j(E_j+E_k)}-{e^{-i E_jt}\over E_jE_k}+{e^{-i(E_j+E_k)t}\over E_k(E_j+E_k)}\right )
 \label{ff2}
\eea

\b

(g) In this manner we can compute any term of any order in $\epsilon$. What we wish to have, however, is a closed form expression for the amplitudes, which we can analyze to see how the probability of excitation moves from the stage $n=0$ to higher stages $n>0$. To do this we will set up a recursion relation below, and solve it to get the various amplitudes. Finally, we will compute the total norm at any stage $n$.

\subsection{Outline of the scheme}

Let us begin by introducing a simple notation. Suppose we are computing $f^{(n)}_{k_1, \dots k_m}(t)$. This gets a contribution from two sources:

\b

(i) From $f^{(n-1)}_{k_1, \dots k_{n-1}}$, through the contribution
\be
{d\over dt}f^{(n)}_{k_1, \dots k_n}(t)= -i\epsilon f^{(n-1)}_{k_1, \dots k_{n-1}}(t)+\dots
\ee
Since the amplitude at stage $n-1$ influences the amplitude at stage $n$, we denote such a contribution by a right arrow $\rightarrow$. 

\b

(ii) From $f^{(n+1)}_{k_1, \dots k_n, k_{n+1}}$, through the contribution
\be
{d\over dt}f^{(n)}_{k_1, \dots k_n}(t)= -i\epsilon^* f^{(n+1)}_{k_1, \dots k_n, k_{n+1}}(t)+\dots
\ee
Since the amplitude at stage $n+1$ influences the amplitude at stage $n$, we denote such a contribution by a left arrow $\leftarrow$. 

 \b
 
 Thus $f^{(2),0}$ computed in (\ref{ff}) is described as $\rightarrow\leftarrow$, while $f^{(2),2}_{jk}$ in (\ref{ff2}) is $\rightarrow\rightarrow$.  Putting together the contributions (i) and (ii) above, we find the equation to be satisfied by the amplitudes
 \bea
 \dot f^{(n)}_{k_1,\dots k_n}&=&-i(E_{k_1}+\dots + E_{k_n})f^{(n)}_{k_1,\dots k_n}-i\epsilon f^{(n-1)}_{k_1, \dots k_{n-1}}-i\epsilon^*\sum_{k_{n+1}}f^{(n+1)}_{k_1, \dots k_n k_{n+1}}\nn
 &\equiv & T_1+T_2+T_3
 \label{eq}
 \eea
 where we have gives names to the three terms on the RHS for later use.
 
 Now let us make some observations about the general pattern of amplitudes.
 
 \subsection{The amplitudes at leading order in $\epsilon$}
 
Consider the amplitude $f^{(n)}_{k_1\dots k_n}$ at any stage $n$. It is easy to see that this amplitude starts at order $\epsilon^n$. This leading order term is given by the symbol
$\rightarrow\rightarrow\dots\rightarrow$. Let us compute these amplitudes. We have for $n\ge 1$
\be
\dot f^{(n),n}_{k_1 \dots k_n}=-i (E_{k_1}+\dots E_{k_n})f^{(n),n}_{k_1\dots k_n}-i\epsilon f^{(n-1),n-1}_{k_1\dots k_{n-1}}
\ee
Writing this as
\be
e^{iE_kt}{d\over dt}\left ( e^{-iE_kt}f^{(n),n}_{k_1 \dots k_n}\right ) = -i\epsilon f^{(n-1),n-1}_{k_1\dots k_{n-1}}
\ee
we get
\be
f^{(n),n}_{k_1 \dots k_n}(t)=(-i\epsilon)e^{iE_kt}\int_{t_1=0}^t dt_1e^{-iE_kt_1}f^{(n-1),n-1}_{k_1\dots k_{n-1}}(t_1)
\label{one}
\ee
Recaling that
\be
f^{(0),0}=1
\ee
we find, using (\ref{one}) recursively
\bea
&&f^{(n),n}_{k_1 \dots k_n}(t)=(-i\epsilon)^ne^{-i(E_{k_1}+\dots + E_{k_n})t}\nn
&&\times\int_{t_n=0}^t dt_n e^{i E_{k_n} t_n}\int_{t_{n-1}=0}^{t_n} dt_{n-1} e^{i E_{k_{n-1}} t_{n-1}}\int_{t_{n-2}=0}^{t_{n-1}} dt_{n-2} e^{i E_{k_{n-2}} t_{n-2}}\dots \int_{t_1=0}^{t_2} dt_1 e^{i E_{k_1} t_1}\nn
\eea
Let us write
\bea
&&I_n(E_{k_1}, \dots E_{k_n},t)=\nn
&&\int_{t_n=0}^t dt_n e^{i E_{k_n} t_n}\int_{t_{n-1}=0}^{t_n} dt_{n-1} e^{i E_{k_{n-1}} t_{n-1}}\int_{t_{n-2}=0}^{t_{n-1}} dt_{n-2} e^{i E_{k_{n-2}} t_{n-2}}\dots \int_{t_1=0}^{t_2} dt_1 e^{i E_{k_1} t_1}\nn
\label{ii}
\eea
Integrating by parts, we find the recursion
\bea
I_n(E_1, \dots E_n, t)&=& {e^{i E_{k_n} t}\over i E_{k_n}}\int_{t_{n-1}=0}^{t} dt_{n-1} e^{i E_{k_{n-1}} t_{n-1}}\int_{t_{n-2}=0}^{t_{n-1}} dt_{n-2} e^{i E_{k_{n-2}} t_{n-2}}\dots \int_{t_1=0}^{t_2} dt_1 e^{i E_{k_1} t_1}\nn
&&-\int_{t_n=0}^t dt_n {e^{i E_{k_n} t_n}\over i E_{k_n}} e^{i E_{k_{n-1}} t_{n}}\int_{t_{n-2}=0}^{t_{n}}dt_{n-2}e^{i E_{k_{n-2}} t_{n-2}}\dots \int_{t_1=0}^{t_2} dt_1 e^{i E_{k_1} t_1}\nn
&=&{e^{i E_{k_n} t}\over i E_{k_n}}I_{n-1}(E_1, \dots E_{k_{n-2}}, E_{k_{n-1}}, t)-{1\over i E_{k_n}}I_{n-1}(E_{k_1}, \dots E_{k_{n-2}},E_{k_{n-1}}+E_{k_n}, t)\nn
\eea

\subsection{Higher orders in $\epsilon$}

We had computed some examples of the amplitudes $f^{(n),m}_{k_1\dots k_n}$ in section \ref{setup}. We can compute more examples the same way, and we find at stage $n=0$
\be
f^{(0),0}=1, ~~~f^{(0),2}=-{|\epsilon|^2\pi \over \Delta} t, ~~~f^{(0),4}={|\epsilon|^2 \pi^2\over 2\Delta^2} t^2
\ee
while the odd orders $f^{(1)}, f^{(3)}, \dots $ vanish. This pattern suggests the ansatz
\be
f^{(0)}=e^{-{|\epsilon|^2\pi\over \Delta} t}
\ee
Checking other cases, we find a similar pattern, which suggests the ansatz
\bea
f^{(n)}_{k_1 \dots k_n}(t)&=&f^{(n),n}_{k_1 \dots k_n}(t)e^{-{|\epsilon|^2\pi\over \Delta} t}\nn
&=&(-i\epsilon)^ne^{-i(E_{k_1}+\dots + E_{k_n})t}I_n(E_{k_1}, \dots E_{k_n},t)e^{-{|\epsilon|^2\pi\over \Delta} t}\nn
\label{ansatz}
\eea
We now verify this ansatz by checking that it satisfies (\ref{eq}). This ansatz gives
\bea
\dot f^{(n)}_{k_1 \dots k_n}(t)&=&(-i\epsilon)^n \left ( -i(E_{k_1}+\dots +E_{k_n})\right ) e^{-i(E_{k_1}+\dots + E_{k_n})t}I_n(E_{k_1}, \dots E_{k_n},t)e^{-{|\epsilon|^2\pi\over \Delta} t}\nn
&+&(-i\epsilon)^ne^{-i(E_{k_1}+\dots + E_{k_n})t}\dot I_n(E_{k_1}, \dots E_{k_n},t)e^{-{|\epsilon|^2\pi\over \Delta} t}\nn
&+&(-i\epsilon)^ne^{-i(E_{k_1}+\dots + E_{k_n})t}I_n(E_{k_1}, \dots E_{k_n},t)\left ( -{|\epsilon|^2 \pi \over \Delta}\right )e^{-{|\epsilon|^2\pi\over \Delta} t}\nn
&\equiv& T'_1+T'_2+T'_3
\eea
We will now check that these terms $T'_1, T'_2, T'_3$ agree with the corresponding terms $T_1, T_2, T_3$ in (\ref{eq}). 

\b

(i) We immediately see that $T'_1=T_1$.

\b

(ii) We note that
\bea
\dot I_n(E_{k_1}, \dots E_{k_n},t)&=&e^{iE_{k_n} t}\int_{t_{n-1}=0}^{t} dt_{n-1} e^{i E_{k_{n-1}} t_{n-1}}\int_{t_{n-2}=0}^{t_{n-1}} dt_{n-2} e^{i E_{k_{n-2}} t_{n-2}}\dots \int_{t_1=0}^{t_2} dt_1 e^{i E_{k_1} t_1}\nn
&=&e^{iE_{k_n} t}I_{n-1}(E_{k_1}, \dots E_{k_{n-1}},t)
\eea
Then we have
\bea
T'_2&=&(-i\epsilon)^ne^{-i(E_{k_1}+\dots + E_{k_n})t}\dot I_n(E_{k_1}, \dots E_{k_n},t)e^{-{|\epsilon|^2\pi\over \Delta} t}\nn
&=&(-i\epsilon)^ne^{-i(E_{k_1}+\dots + E_{k_n})t}e^{iE_{k_n} t}I_{n-1}(E_{k_1}, \dots E_{k_{n-1}},t)e^{-{|\epsilon|^2\pi\over \Delta} t}\nn
&=&(-i\epsilon)(-i\epsilon)^{n-1}e^{-i(E_{k_1}+\dots + E_{k_{n-1}})t}I_{n-1}(E_{k_1}, \dots E_{k_{n-1}},t)e^{-{|\epsilon|^2\pi\over \Delta} t}\nn
&=&(-i\epsilon) f^{(n-1)}_{k_1, \dots k_{n-1}} = T_2
\eea

\b

(iii) Let us start from $T_3$ in (\ref{eq}) and perform the sum over $k_{n+1}$ using (\ref{sum})
\bea
T_3&=&-i\epsilon^*\sum_{k_{n+1}}f^{(n+1)}_{k_1, \dots k_n k_{n+1}}\nn
&=&-i\epsilon^*{1\over \Delta}\int_{-\infty}^\infty dE_{k_{n+1}}(-i\epsilon)^{n+1}e^{-i(E_{k_1}+\dots + E_{k_n}+E_{k_{n+1}})t}I_{n+1}(E_{k_1}, \dots E_{k_n},E_{k_{n+1}},t)e^{-{|\epsilon|^2\pi\over \Delta} t}\nn
&=&-i|\epsilon|^2{(-i)\over \Delta}(-i\epsilon)^n\int_{-\infty}^\infty dE_{k_{n+1}} e^{-i(E_{k_1}+\dots + E_{k_n})t}e^{-i E_{k_{n+1}}t}\nn
& & \times~~
\left ( {e^{iE_{k_{n+1}} t}\over iE_{k_{n+1}}}I_n(E_{k_1}, \dots E_{k_n},t)-{1\over i E_{k_{n+1}}}I_n(E_{k_1}, \dots E_{k_n}+E_{k_{n+1}},t)\right ) e^{-{|\epsilon|^2\pi\over \Delta} t}\nn
&=&-i|\epsilon|^2{(-i)\over \Delta}(-i\epsilon)^n e^{-i(E_{k_1}+\dots + E_{k_n})t}e^{-{|\epsilon|^2\pi\over \Delta} t}\nn
& & \times~~
\int_{-\infty}^\infty dE_{k_{n+1}} \left ( {1\over iE_{k_{n+1}}}I_n(E_{k_1}, \dots E_{k_n},t)-{e^{-i E_{k_{n+1}}t}\over i E_{k_{n+1}}}I_n(E_{k_1}, \dots E_{k_n}+E_{k_{n+1}},t)\right ) \nn
\eea
We can evaluate the integral
\be
K\equiv \int_{-\infty}^\infty dE_{k_{n+1}} \left ( {1\over iE_{k_{n+1}}}I_n(E_{k_1}, \dots E_{k_n},t)-{e^{-i E_{k_{n+1}}t}\over i E_{k_{n+1}}}I_n(E_{k_1}, \dots E_{k_n}+E_{k_{n+1}},t)\right )
\label{kk}
\ee 
by methods similar to those described in section \ref{setup}, step (e):

\b

(i) We first note that the integrand is smooth on the entire real line; the apparent singularity caused by the factor ${1\over E_{k_{n+1}}}$ cancels between the two terms in the integrand. Thus we can deform the $E_{k_{n+1}}$ contour in a manner similar to fig.\ref{fcontour}, so that it passes below the pole $E_{k_{n+1}}=0$.

\b

(ii) We next close the contour $-\infty<E_{k_{n+1}}<\infty$ by adding a semicircle at infinity in the lower half plane. The integrand is analytic inside this closed contour,  contour, so the integral over the resulting closed contour is zero. But the integral over the semicircle at infinity is not zero, so we have to subtract its value $C$. To compute $C$, we note that the second term in the integrand in (\ref{kk}) vanishes since $e^{-i E_{k_{n+1}}t}\r 0$ on the semicircle; this follows since we are looking at the evolution for $t>0$. But the first term gives a contribution to $K$ equalling
\be
C=-\pi i  {1\over i}I_n(E_{k_1}, \dots E_{k_n},t)=-\pi I_n(E_{k_1}, \dots E_{k_n},t)
\ee
Subtracting this value, we find that
\be
K=\pi I_n(E_{k_1}, \dots E_{k_n},t)
\ee
and
\bea
T_3&=&-i|\epsilon|^2{(-i)\over \Delta}(-i\epsilon)^n e^{-i(E_{k_1}+\dots + E_{k_n})t}e^{-{|\epsilon|^2\pi\over \Delta} t}~ \pi I_n(E_{k_1}, \dots E_{k_n},t) \nn
&=& T'_3
\eea

\b

We thus verify that the ansatz  (\ref{ansatz}) indeed solves the equations (\ref{eq}), and thus gives a solution to the problem to all orders in $\epsilon$.

\subsection{Computing norms}

From (\ref{ansatz}), we see that the amplitude in stage $n=0$ at time $t$ is 
\be
f^{(0)}(t)=e^{-{|\epsilon|^2\pi\over \Delta} t}
\ee
Thus the probability in the stage $n=0$ is
\be
P^{(0)}(t)=\left | f^{(0)}(t) \right |^2=e^{-{2|\epsilon|^2\pi\over \Delta} t}
\ee
This decays with time as expected.

We would now like to compute the probability $P^{(n)}(t)$ in all stages $n$. We have
\be
P^{(n)}(t)=\sum_{k_n}\dots \sum_{k_1} \left | f^{(n)}_{k_1, \dots k_n}(t) \right |^2
\ee
Writing the sums as integrals  and using (\ref{ansatz}) gives
\be
P^{(n)}(t)={|\epsilon|^{2n}\over \Delta^n}e^{-{2|\epsilon|^2\pi\over \Delta} t}\int_{-\infty}^\infty dE_{k_n}\dots \int_{-\infty}^\infty dE_{k_1} ~ I_n(E_{k_1}, \dots E_{k_n},t)I^*_n(E_{k_1}, \dots E_{k_n},t)
\label{pp}
\ee
Let us therefore compute
\be
L_n(t)=\int_{-\infty}^\infty dE_{k_n}\dots \int_{-\infty}^\infty dE_{k_1} ~ I_n(E_{k_1}, \dots E_{k_n},t)I^*_n(E_{k_1}, \dots E_{k_n},t)
\ee
Noting the definition (\ref{ii}) of $I_n$ we have
\bea
L_n(t)&=&\int_{-\infty}^\infty dE_{k_n}\dots \int_{-\infty}^\infty dE_{k_1} ~\nn
&&\times \left ( \int_{t_n=0}^t dt_n e^{i E_{k_n} t_n}\int_{t_{n-1}=0}^{t_n} dt_{n-1} e^{i E_{k_{n-1}} t_{n-1}}\dots \int_{t_1=0}^{t_2} dt_1 e^{i E_{k_1} t_1}\right )\nn
&&\times \left ( \int_{t'_n=0}^t dt'_n e^{-i E_{k_n} t'_n}\int_{t'_{n-1}=0}^{t'_n} dt'_{n-1} e^{-i E_{k_{n-1}} t'_{n-1}}\dots \int_{t'_1=0}^{t'_2} dt'_1 e^{-i E_{k_1} t'_1}\right )\nn
\label{ll2}
\eea
Let us first perform the integral over $E_{k_n}$. Interchanging the orders of some integrals, we get
\bea
L_n(t)&=&\int_{-\infty}^\infty dE_{k_{n-1}}\dots \int_{-\infty}^\infty dE_{k_1} \left (\int_{t_n=0}^t dt_n
 \int_{t'_n=0}^t dt'_n \int_{-\infty}^\infty dE_{k_n}e^{i E_{k_n} (t_n-t'_n)}\right ) ~\nn
&&\times \left ( \int_{t_{n-1}=0}^{t_n} dt_{n-1} e^{i E_{k_{n-1}} t_{n-1}}\dots \int_{t_1=0}^{t_2} dt_1 e^{i E_{k_1} t_1}\right )\nn
&&\times \left (\int_{t'_{n-1}=0}^{t'_n} dt'_{n-1} e^{-i E_{k_{n-1}} t'_{n-1}}\dots \int_{t'_1=0}^{t'_2} dt'_1 e^{-i E_{k_1} t'_1}\right )\nn
\label{l}
\eea
We have
\bea
\left (\int_{t_n=0}^t dt_n
 \int_{t'_n=0}^t dt'_n \int_{-\infty}^\infty dE_{k_n}e^{i E_{k_n} (t_n-t'_n)}\right ) &=&
 \left (\int_{t_n=0}^t dt_n
 \int_{t'_n=0}^t dt'_n ~~2\pi \, \delta (t_n-t'_n)\right ) \nn
 &=&  \left (\int_{t_n=0}^t dt_n
 ~~2\pi \right ) \nn
 \label{lpart}
  \eea
where now $t'_n=t_n$. Putting this in (\ref{l}) gives
\bea
L_n(t)&=&2\pi \int_{t_n=0}^t dt_n
\int_{-\infty}^\infty dE_{k_{n-1}}\dots \int_{-\infty}^\infty dE_{k_1}   ~\nn
&&\times \left ( \int_{t_{n-1}=0}^{t_n} dt_{n-1} e^{i E_{k_{n-1}} t_{n-1}}\dots \int_{t_1=0}^{t_2} dt_1 e^{i E_{k_1} t_1}\right )\nn
&&\times \left (\int_{t'_{n-1}=0}^{t_n} dt'_{n-1} e^{-i E_{k_{n-1}} t'_{n-1}}\dots \int_{t'_1=0}^{t'_2} dt'_1 e^{-i E_{k_1} t'_1}\right )\nn
&=&2\pi \int_{t_n=0}^t dt_n \, L_{n-1}(t_n)
\label{ll}
\eea
From (\ref{l}),(\ref{lpart}) we see that
\be
L_1(t)= \left (\int_{t_1=0}^t dt_1 ~~2\pi \right ) =2\pi t
\ee
Then using (\ref{ll}) recursively, we find
\bea
L_n(t)&=&2\pi \int_{t_n=0}^t dt_n \, L_{n-1}(t_n)\nn
&=&(2\pi )^2 \int_{t_n=0}^t dt_n  \int_{t_{n-1}=0}^{t_n} dt_{n-1} \, L_{n-2}(t_{n-1})\nn
&=&(2\pi)^{n-1} \int_{t_n=0}^t dt_n  \int_{t_{n-1}=0}^{t_n} dt_{n-1}\dots \int_{t_2=0}^{t_3} L_1(t_2)\nn
&=&(2\pi)^{n} \int_{t_n=0}^t dt_n  \int_{t_{n-1}=0}^{t_n} dt_{n-1}\dots \int_{t_2=0}^{t_3} t_2\nn
&=&{(2\pi)^n t^n\over n!}
\eea
Substituting this in (\ref{pp}) we finally get
\be
P^{(n)}(t)={1\over n!}\left ( {2\pi |\epsilon|^2 t\over \Delta }\right ) ^n e^{- {2\pi |\epsilon|^2 t\over \Delta }}
\label{pq}
\ee

\subsection{The peak of the norm}

At $t=0$ the norm $P^{(n)}(t)$ is entirely peaked at the stage $n=0$, and at later times it spreads to larger values of $n$. For late times $t$, we wish to find the value of $n$ where this norm is peaked. For this purpose we treat $n$ as a continuous variable, and maximize $P^{(n)}(t)$ over $n$. We have to maximize a function of the form
\be
h(x)={x^n\over n!} e^{-x}
\ee
where in our case $x={2\pi |\epsilon|^2 t\over \Delta }$. 
We have
\be
{dh\over dx}=e^{-x}\left ({x^{n-1}\over (n-1)!}-{x^n\over n!}\right )=0
\ee
The peak of $h$ corresponds to the vanishing of this derivative, which occurs at
\be
x=n
\ee
Thus we find that the peak of the norm occurs at the stage
\be
n(t)={2\pi |\epsilon|^2 t\over \Delta }
\label{peak}
\ee
Thus with the quantum mechanical problem that we have set up, the norm moves along the successive stages at a constant velocity.

\section{Breakdown of the principle of equivalence}

There are two aspects of black hole physics that we now wish to discuss. The first, which we will do in this section, addresses how the principle of equivalence can break down at the horizon of a black hole. The semiclassical approximation would naively appear to be valid for infall through a horizon,  but as we will see, this is not the case in a theory with microstates like fuzzballs which have structure at the horizon. The second aspect is the idea of fuzzball complementarity, for which we have made a model in the above sections; in the next section we will put together this model with more abstract ideas to present a more complete picture of this complementarity.

\subsection{The traditional principle of equivalence}

We study the electromagnetic, weak and strong interactions using field theory in flat space. But we think of gravity as being a theory is curved space. Why do we have this difference?

At one level we should be able to do gravity using flat space; after all the graviton is a gauge boson like the photon, and if we look at Feynman diagrams in flat space then we should get  the effects of gravity. There should be an {\it equivalent} way to get the same effects through a modification of space to include curvature. The reason we normally think of gravity in the latter way rather than the former is that we  cannot be sure that the field theory approach will capture situations that are not asymptotically flat, like our Cosmology, or the  effects of nontrivial topology like wormholes.

But as we will see now, even for black holes the relation between the two issues needs to be examined carefully, which is what we will proceed to do now.

\begin{figure}[h]
\begin{center}
 \includegraphics[scale=.58] {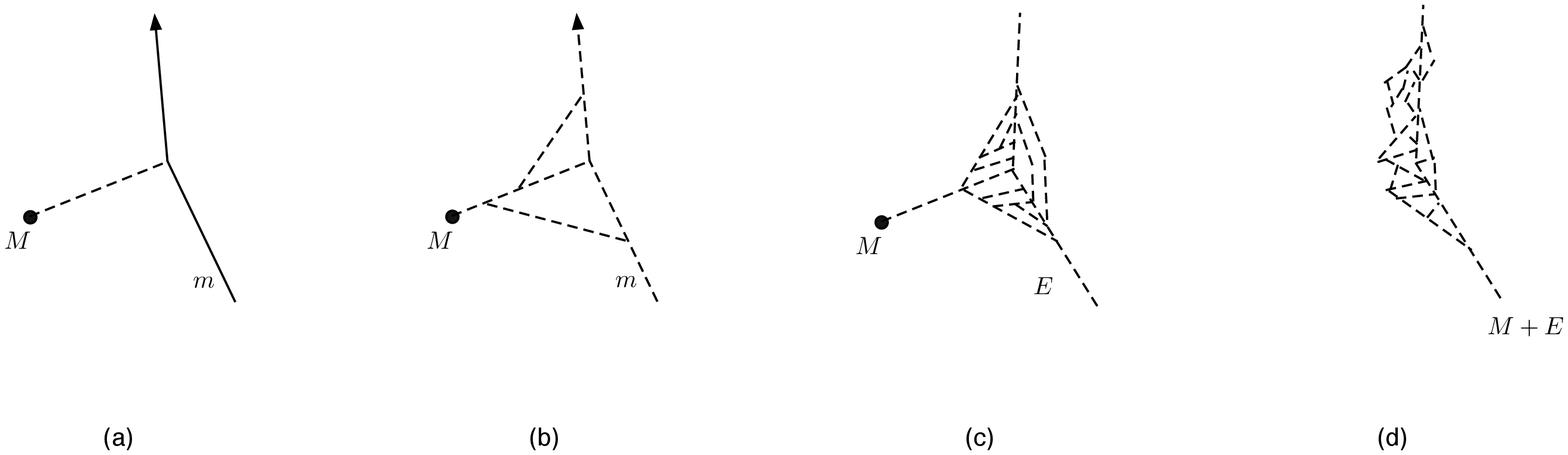}
\caption{\label{fbone} (a) A graviton exchange, using flat spacetime (b) The nonlinearity of gravity gives a complex of graviton propagators (c) At the threshold of horizon formation, the interaction between this complex of gravitons makes the entire evolution freeze in a theory without fuzzballs (d) If there are alternative gravity solutions available (the fuzzballs), then the evolution does not freeze, but instead continues into the directions of superspace described by the  fuzzballs}
\end{center}
\end{figure}

How do we relate the field theory in flat space to the idea of no forces in curved space? This is of course well known, and a discussion can be found in \cite{feynman}. Let us start with the weak field approximation, where one particle (mass $m$) scatters off the gravitational field of a second particle (mass $M$). The leading order Feynman diagram for this is given in fig.\ref{fbone}(a). The free action for the particle can be written (in a certain gauge) as
\be
-m\int d\tau  \h {d x^\mu\over d\tau}{d x^\nu\over d\tau}\eta_{\mu\nu} 
\label{free}
\ee
The gravitational coupling has the form
\be
 -\lambda h_{\mu\nu}T^{\mu\nu}
\label{couplinggrav}
\ee
with
\be
T^{\mu\nu}=m U^\mu U^\nu, ~~~U^\mu={dx^\mu\over d\tau}
\ee
 The  particle $M$ produces the graviton $h_{\mu\nu}$ through the coupling (\ref{couplinggrav}). The  particle $m$ then gets a total action  
\be
S=-m\int d\tau \h {dx^\mu\over d\tau}{dx^\nu\over d\tau}(\eta_{\mu\nu}+ 2\lambda h_{\mu\nu})
\ee
Now defining a new metric
\be
g_{\mu\nu}\equiv \eta_{\mu\nu}+ 2\lambda h_{\mu\nu}
\label{newg}
\ee
we get the dynamics of the particle in a form where we have no interactions but a new metric
\be
S=-m\int d\tau \h {dx^\mu\over d\tau}{dx^\nu\over d\tau}g_{\mu\nu}
\ee
This is the way we generate curved space; the `stretching'  of the space under gravity is a rewriting of the Feynman diagrams giving graviton exchange. 

But gravity is a nonlinear theory, so gravitons themselves produce other gravitons. We depict this in fig.\ref{fbone}(b). As the gravitational field of $M$ becomes stronger, we get a more and more dense set of gravitons lines. One effect of these interactions is redshift: everything slows down. Thus we may say that repeated interactions between the gravitons leads to a situation where the state is not able to evolve any further; this corresponds to $m$ reaching the horizon radius generated by $M$.

What happens then? In the standard treatment of the black hole, we abandon the field theoretic approach at this point, and proceed with the curved space metric $g_{\mu\nu}$ as an entity that has included all interactions between $m$ and $M$. We then note that a change to Kruskal coordinates allows us to continue $g_{\mu\nu}$ to a region interior to the horizon, with no singularity at the horizon. It therefore appears that nothing really happens at the horizon.

We will now see that in a theory with fuzzballs such a conclusion does not follow. We will present a picture, first in the gravity description and then in the dual CFT, which shows an alternate possibility where the system evolves to a linear combination of fuzzballs when $m$ reaches the horizon.

In a unified theory of gravity like string theory, the particle $m$ is not really distinct from the fields of the gravitational theory. Thus we can think of $m$ as being replaced by a graviton. Then the evolution of fig.\ref{fbone}(b) can be thought of as just the evolution of a complicated gravitational excitation, in a domain where gravity has become very nonlinear. In fact the AdS/CFT duality \cite{maldacena} encourages us to think of gravity in this way.

\subsection{The gravity picture of the evolution}

The particle that is falling in the field of $M$ is not a test particle; we assume that it has a nonzero energy $E$. Because this particle itself creates a gravitational field, we have additional graviton propagators as shown in fig.\ref{fbone}(c). It is under the combined effect of the propagators arising from $M$, and the propagators arising from the infalling particle, that the complex of graviton lines becomes so dense that evolution freezes. This dependence on $E$  corresponds to the fact that the horizon created by $M$ moves out to a location depending on $E$.

If we just took canonically quantized gravity, which has no fuzzball states, then this would be the end of the story; the evolution freezes  at the horizon in Schwarzschild coordinates, and we can continue further inwards  in Kruskal coordinates. But in a theory with fuzzballs, the complex of gravitons starts to evolve into a large vector space of new configurations -- the fuzzball states $|F_i\rangle$. Fuzzball states are states of the gravity theory with structure at the horizon. It was argued in \cite{tunnel} that the large number of fuzzball states creates an order unity probability for the incoming particle wavefunction to move into the space of fuzzball states. We depict this evolution in fig.\ref{fbone}(d) where the gravitons have become deformed into new configurations.  

Now we do not just have a frozen evolution in the Schwarzschild frame; rather, the evolution continues along new directions in superspace -- the space of all gravity solutions.  In string theory this superspace contains all the fuzzball solutions which have no horizon, but instead have a nontrivial structure at the place where the horizon would have appeared. 

\subsection{The CFT picture}

In a set of papers \cite{cft} the evolution of perturbations was studied in the dual CFT. While these computations address only very low point correlators, we will use the general picture suggested by these computations to extrapolate to the situation where black holes form.

Suppose we drop a graviton  into one of the  ground states of the D1D5 system. The typical  ground state has a high winding number for the component strings, and we assume that such is the case. The infalling graviton will, in the classical picture, make a black hole horizon after it falls an appropriate distance down the throat. We will now track the evolution in the CFT and see that the evolution can in fact reproduce the alternative picture discussed above where we get a fuzzball state at the horizon in place of the vacuum of the traditional hole. 

The initial state of the graviton is created by an operator of the form \cite{comparing}
\be
\h (\p X^i\bar \p X^j+\p X^j\bar \p X^i)
\ee
Thus, roughly speaking, we create one pair of excitations (one left moving excitation and one right moving excitation). 

The interaction in the CFT is a `twist operator' plus a supercharge.  This operator can generate additional excitations, subject only to the constraint of overall energy and momentum conservation. As the graviton falls deeper down the throat, it reaches locations that correspond to lower energies in the CFT. But the energy of the graviton is a conserved quantity. Thus the initial pair of excitations break up into  a larger number of excitations, each with lower energy. As the graviton keeps falling down the throat, its CFT dual  becomes a cloud of more and more excitations, each with an energy such that the overall energy is still the energy of the initial graviton. 

Lower energy excitations lave a longer wavelength, and so occupy more space on the `effective string', which had a total length $n_1n_5 L$. The fact that we have more excitations also contributes to the fact that  a larger part of this string will be  occupied. It was found in \cite{lm4} that when the excitations cover a fraction of order unity on the effective string, then in the gravity picture the infalling graviton is at the threshold of making  a black hole. (In \cite{lm4} the effective strong was taken to be broken into several loops, but this choice is irrelavent; one could consider a single multiwound effective string.) This is the CFT analogue of black hole formation: the infalling graviton needed more and more space on the effective string, but at some point the effective string runs out of space, and this natural evolution cannot proceed further. We wish to ask: what happens then? Can we see something here that mirrors what we conjectured above for the gravity description?

\subsubsection{Redshift in the CFT}

In the gravity description the principal feature of approaching a horizon is the divergence of the redshift. Let us ask how this is seen in the CFT. The redshift is a very universal effect in gravity, acting equally on all particles at a give location, so it must have an equally universal description in the CFT. The dual of the graviton is the stress tensor $T$, so let us look at the effects generated by $T$.

Recall that the CFT has $n_1n_5$ copies of a basic $c=6$ CFT. In the free CFT, the stress tensor has the form
\be
T=\sum_{i=1}^{n_1n_5} \p X^{(i),k}\p X^{(i),k}
\label{stress}
\ee
where $X^{(i),k}$ is the $k$ component of $X$ in the copy $i$ of the CFT, and $k$ is assumed summed as $k=1, \dots 4$.  At this free level, we have many operators of the same dimension $\Delta=2$; for example
\be
{\cal O}=\p X^{(1),k}\p X^{(1),k}-\p X^{(2),k}\p X^{(2),k}
\label{calo}
\ee
But when we turn on the interaction,   the gravity dual implies that  the graviton multiplet will be the only massless field. Thus the diagonal combination (\ref{stress}) will still have $\Delta=2$, while all the other combinations will be lifted to very high dimensions. This lifting leads to a direct coupling between the different copies. Suppose for simplicity that there are only two copies. Now suppose we have an excitation on copy $1$ which generates a stress tensor
\be
\p  X^{(1),k}\p X^{(1),k}
 \ee
 We can write this as
 \be
 \h \left ( \p X^{(1),k}\p X^{(1),k}+\p X^{(2),k}\p X^{(2),k}\right ) +  \h \left ( \p X^{(1),k}\p X^{(1),k}-\p X^{(2),k}\p X^{(2),k}\right )
\ee
Since the operator in the second bracket is lifted to a high dimension, we effectively get only the operator in the first bracket. (The contributions of the operator in the second bracket phase cancel in low energy processes.) Thus on copy $2$ we find an effective operator
\be
\h \p X^{(2),k}\p X^{(2),k}
\label{teff}
\ee
Thus we can say (in our rough qualitative picture) that  an excitation on copy $1$ generates a pair of $X$ excitations on copy $2$.

Suppose we start with a large number ${ N}$ of copies. We are interested in the behavior of an excitation on copy $1$. But the above interaction through the stress tensor couples the excitations on copy $1$ to the excitations on all the other copies. At the threshold of black hole formation, the entire effective string is occupied by excitations, so the effect of this coupling is very important. Let us see if we can extract a simple consequence of this interaction.

Since we are interested in copy $1$, let us try to integrate out the other copies one by one. Start with copy $N$, and consider its effect on copies $i=1, \dots N-1$. By an interaction like (\ref{teff}), we get an effective stress tensor $T$ on these copies $i=1, \dots N-1$. Suppose that the excitations are at a wavelength $\lambda_0$. Then for wavelengths $\lambda\gtrsim \lambda_0$, we average over excitations like (\ref{teff}), getting an expectation value for
\be
T+\bar T ~\r~ \langle T \rangle + \langle \bar T \rangle
\label{replace}
\ee
Integrating over all values of the spatial coordinate $\sigma$ gives
\be
\int_{\sigma=0}^{2\pi} d\sigma \left ( \langle T \rangle + \langle \bar T \rangle \right ) =H
\ee
where $H$ is the Hamiltonian. Since the Hamiltonian generates evolution as $e^{-iHt}$, we get
\be
 H ~\r~ i\p_t
 \ee
 This is of course just the conformal Ward identity. It tells us that we can replace the effect of $T$ by an increase in the length of the time direction of the cylinder on which the 1+1 dimensional CFT is defined
 \be
 t\r (1+a) t
 \ee
 for some $a>0$; the sign of $a$ is positive since energy density $\langle T \rangle + \langle \bar T \rangle$ is a positive quantity. 
 Now we integrate out a the copy $N-1$, and examine the effect on the remaining copies $i=1, \dots N-2$. Again we get an increase in the effective $t$ coordinate for these copies, so we now have
 \be
 t\r (1+a)^2 t
\ee
Proceeding in this way till we have only copy $1$ left, we find 
\be
t\r (1+a)^{N-1}t\approx e^{aN}t
\ee
where in the second step we have assumed that $a\ll 1$, and $N\gg 1$. We now see that for
$aN\ll 1$
we will get the usual time coordinate $t$, while for $aN\gg 1$, we will find a very large slowdown of the time. It is the latter situation which corresponds to horizon formation in the CFT.\footnote{We have ignored the fact that each time we integrate out one copy, the state on the remaining copies changes a little.  More explicit computations can done by taking specific choices of operators and states, but all we wish to note here is that the redshift diverges for low energy modes (those for which the replacement (\ref{replace}) can be made)  when $N\r\infty$, for situations where the effective string is fully covered by excitations. In alternate ways of estimating the redshift, we can get the redshift to behave as $1/(1-aN)$; but this is pointing to the same physics.}

\subsubsection{Evolution after reaching the horizon}

Thus we see that in the CFT something does happen at the horizon: 

\b

(1)  The effective string fills up with excitations \cite{lm4}.

\b

(2) There is a large slowdown in the normal evolution of states; this is the analogue of the classical approach to horizon formation. What we will now see is that there is a  further evolution possible for the CFT state which will mirror the evolution of the gravity state to fuzzballs,  which was depicted in fig.\ref{fbone}(d). 

\b

If we have a few excitations on the effective string, then operators like (\ref{calo}) are lifted to high dimensions $\Delta$, and so do not contribute to the dynamics. But when the effective string is completely filled up, a large class of new low dimension operators appears. This is of course just due to deconfinement, an effect that can be found in nuclear physics. At low densities, quarks join up into hadrons because of confinement, but if the density is high enough, then a quark can find a `partner' somewhere nearby in the plasma, so the entire state does not split up into well separated hadrons. Black hole formation in the CFT is expected to correspond to a deconfinement type transition \cite{wittenthermal}, in agreement with the picture we are noting here. Support for this picture comes from the fact that the free CFT reproduces the entropy of the near extremal black hole \cite{callanmalda}; if most of the excitations in the CFT were lifted to high dimensions, we would get  insufficient entropy to reproduce the black hole value. 

In the CFT we have seen that we do get a slowdown in evolution which is the analogue of redshift, but there is a whole new set of directions where the CFT state can evolve; these are the deconfined states which are available only when the density of excitations on the effective string becomes high enough so that {\it all} the effective string is occupied. Thus the evolution does not have to freeze; the wavefunction spreads along the large space of deconfined states. 

But now we should ask: what are the analogues in gravity of these deconfined states in the gravity description? We expect that these are the fuzzball states $|F_i\rangle$. Prior to the discovery of the fuzzball construction, it was believed that the horizon had to be a  smooth place, where the evolution of low energy modes would be just like evolution in the vacuum. It was then argued that the information paradox would be resolved through small corrections to this vacuum evolution. But in \cite{cern} it was shown that this cannot work; one needs {\it order unity} correction to the evolution of low energy physics at the horizon. The issue then becomes: how do we break the no-hair theorem and produce such order unity corrections? 

This breaking of the no-hair theorem is what the fuzzball construction achieves. In \cite{gibbonswarner} it was shown how the assumptions used in the traditional no-hair theorems fail in string theory. With the existence of fuzzball states, we can now argue that in the gravity picture the wavefunctional starts to spread over the space of these fuzzball states. Thus we get a different evolution from the evolution in  a theory without fuzzballs; in the latter case, the evolution just froze at the horizon, and could be analytically continued to smooth evolution in the interior. But with fuzzballs the evolution has taken a different direction, where the wavefunctional starts spreading over a large space of new states at the horizon. Thus we do not have the normal continuation to infall through a smooth interior. 

But can we recover an {\it effective} interior for special situations? As we have noted above, the conjecture of fuzzball complementarity says that for freely falling objects with $E\gg T$, the evolution of the fuzzball surface can mimic infall through an effective interior. Let us now turn to this issue.

\section{Fuzzball complementarity}

Let us now return to the main theme of this paper: the notion of fuzzball complementarity. In the above section we have seen that the infall of a shell stops at the horizon, and the wavefunctional starts to spread over the space of fuzzball states $|F_i\rangle$. The fuzzball states can be taken as a complete basis of states of the gravity theory, so we can write
\be
|\psi\rangle = \sum_i C_i |F_i\rangle
\ee
The evolution then proceeds as
\be
\sum_i C_i |F_i\rangle\r \sum_i C_i e^{-iE_i t}|F_i\rangle
\label{fc}
\ee
What is crucial to note is that this evolution should be described in superspace -- the space of all solutions in gravity. We are not looking at the dynamics of particles in any one metric. The conjecture of fuzzball complementarity then says that the evolution (\ref{fc}) can be mapped {\it approximately} to the evolution of an infalling shell in the metric of the traditional black hole. The latter evolution is of course something that happens in just one metric -- the metric of the traditional hole. This conjecture was termed gravity - gravity duality in \cite{blackholo}; it says that {\it the evolution in superspace of the full gravity wavefunctional can be mapped, in an approximation valid for situations like black hole formation, to an evolution in a single gravity solution}. 

As we have noted in the above sections, such a conjecture is motivated by the idea of AdS/CFT duality. But the two conjectures are very different in what they try to say:

\b

(a) In AdS/CFT duality, we say that the gravity theory can me mapped, under a very nontrivial change of variables, to a Yang-Mills theory. The map is exact, and relates two different theories to each other. Put another way, the spectrum of gravity in an asymptotically AdS space is the same as the spectrum of Yang-Mills theory on the boundary of this space; thus a change of variables can map one theory to the other. 

\b

(b) In gravity-gravity duality, we consider just the gravitational theory, and not any alternative description of this theory in terms of a field theory. We consider situations like black hole formation where the wavefunctional of this gravity theory spreads over a large superspace. We ask if there is an {\it approximate} description of this evolution in some new variables. We conjecture that there is in fact an approximate duality map between this evolution in superspace and the evolution in  the traditional black hole background. As the full wavefunctional spreads out in superspace, in the approximate dual description we see the infall of a wavefunction in the interior of the traditional hole.

\b

\begin{figure}[h]
\begin{center}
 \includegraphics[scale=.72] {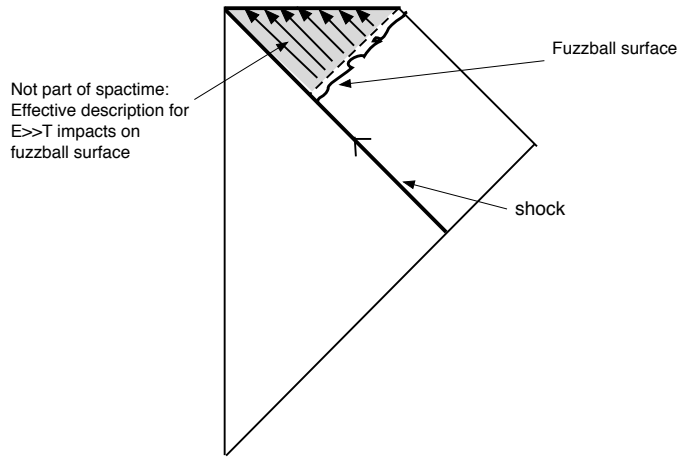}
\caption{\label{penroseshock} A modified notion of a Penrose diagram. The nonshaded part of the diagram is actual spacetime. The shaded part with arrows is an effective approximation from $E\gg T$ quanta that are inward directed; the actual gravity wavefunctional is a linear combination of states that have no horizon.}
\end{center}
\end{figure}

To illustrate the conjecture of fuzzball  complementarity, we draw a Penrose type diagram in fig.\ref{penroseshock}. The incoming solid line with an arrow denotes a collapsing shell. The gravity effects discussed in the last section halt the natural evolution at the place where a horizon would have formed, and the full wavefunctional begins to spread over the nontrivial fuzzball configurations of the theory. This behavior is indicated by the thick wiggly line just outside the horizon. The interior of the horizon arises only in the approximation of fuzzball complementarity, as an approximate description of the evolution at the location of the thick wiggly line. Since this is an approximate, effective dual description, we mark this region with a set of lines. These lines carry inward pointing arrows, indicating that this effective description emerges only for infalling quanta that are falling in hard ($E\gg T$) from the right. Thus this is not a region of normal spacetime; it is an effective spacetime only for a certain kinematic regime of trajectories. In particular this region does not describe right moving modes just inside the horizon, which would have corresponded to the negative energy partners of Hawking radiation. 

\subsection{The loophole in the AMPS argument}

Let us now recall how the idea of fuzzball complementarity bypasses the argument by Almheiri, Marolf, Polchinski and Sully (AMPS) \cite{amps} which says that one cannot get complementarity:

\b

(a) AMPS were looking for an {\it exact} complementary description, not one that emerges an an approximate one  in the limit $E\gg T$ for particles that fall freely from afar onto the black hole. In particular, they consider experiments where an observer swoops down and picks up precisely the $E\sim T$ quanta from near the horizon, and argue that these quanta cannot be given a complementary description where the horizon is in a vacuum state for these modes. But such $E\sim T$ quanta are not covered in the description of fuzzball complementarity.

\b

  \begin{figure}[!]
\includegraphics[scale=.72]{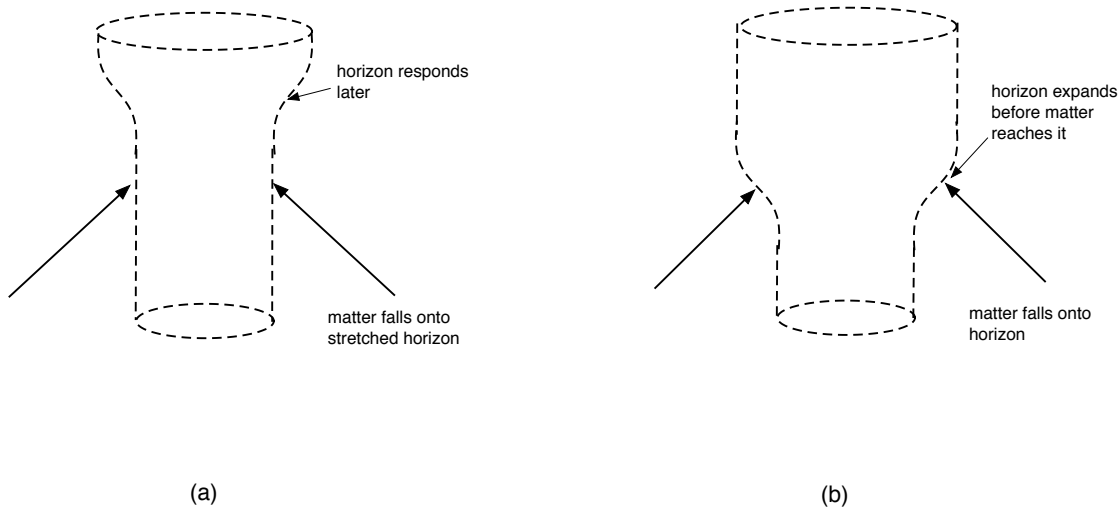}
\caption{ (a) AMPS assumed that the boundary of the hole -- as defined by the `stretched horizon' -- responds only after infalling matter reaches it; this assumption violates the Bekenstein limit. (b) In standard general relativity, it is known that the horizon expands {\it before} matter reaches it; that is why the Bekenstein limit is not violated. }
\label{famps}
\end{figure}

(b) AMPS assumed that the stretched horizon does not respond in any way until an infalling particle hits it (fig.\ref{famps}(a)). But a little reflection shows that there is something odd about this assumption. The stretched horizon is normally expected to lie a planck length {\it outside} the event horizon. But the event horizon expands outwards {\it before} an infalling object reaches it. Thus should it not be the case that the stretched horizon expands outwards before the infalling object hits it (fig.\ref{famps}(b))?

The reason this issue did not come up for AMPS was that they did not have any criterion like $E\gg T$ in their approach to complementarity, so they could replace the infalling object by a test particle of negligible energy, in which case the expansion of the horizon would be ignorable. 

But if we are interested only in high energies $E\gg T$, then we must address the issue of whether the stretched horizon moves out before the particle reaches it. In \cite{mt2} it was argued that the stretched horizon must indeed move out in advance of the particle reaching it; otherwise we violate the Bekenstein limit which says that the maximum entropy in a region is given by $S=A/4$, with $A$ the area of the stretched horizon. This is easy to see, as follows. After the hole has passed its halfway evaporation point, it is maximally entangled with its emitted radiation, and so has the maximal value $A/4$ for the entropy of the stretched horizon. If an infalling particle carrying one bit of information lands on the stretched horizon {\it without there being an earlier expansion of the stretched horizon} then we have $A/4+1$ bits on an area $A$. This is in contradiction with the traditionally assumed entropic property of the stretched horizon.

There is a second, equally difficult problem with the AMPS assumption. If a shell of mass $m$ lands on a stretched horizon with mass $M$, the resulting object has mass $M+m$ and is thus inside its horizon. Then the light cones point inwards at the old stretched horizon, and one needs acausal evolution to make the new stretched horizon move out to its correct location. But AMPS have shown no evidence of how acausal behavior is possible in their theory.

If the stretched horizon indeed expands outwards before the infalling particle reaches it, then the AMPS argument against complementarity does not work. The reason is that Hawking radiation has a very low energy; it reaches planck temperature only when we reach the stretched horizon. If we look at a location that is {\it outside} the stretched horizon by some distance $D$, then the temperature of the Hawking radiation here would drop to a value that becomes smaller as $D$ becomes larger. As we increase the energy $E$ of the infalling particle, the distance $D$ increases, and the temperature of the radiation it feels before reaching the (new) stretched horizon decreases as well. Since the AMPS argument involved the burning up of the infalling particle by the Hawking radiation, this burning effect drops off with larger $E$. As a consequence, the infalling particle with $E\gg T$ can reach the stretched horizon without significant interaction with the emerging Hawking radiation, and one cannot rule out the possibility that the further evolution of the bits on the stretched horizon will give an approximate complementarity.

\b

(c) While the above argument indicates that we  should indeed let the stretched horizon move outwards before it is impacted by the infalling particle, let us also review the reasons why AMPS had assumed that the stretched horizon would not move in this fashion.  The argument was essentially one of causality. The incoming particle could be falling in radially at the speed of light. If signals cannot propagate faster than $c$, then the stretched horizon will not know that the particle is coming in, and so should not respond until it is hit.

Let us make this more precise.  Consider a location that is at a distance $D$ outside the stretched horizon. Suppose this is the position to which the horizon would move out in classical gravity when the incoming quantum falls in. One could argue that when the incoming particle reaches this location, it is still travelling through a locally empty spacetime. Thus it does not `hit' anything, and so should not feel any novel effects. In fact one of the assumptions of AMPS is that there should be no new effects outside the stretched horizon; for example one should not have  nonlocal effects of the kind suggested by Giddings \cite{giddings}. So when the particle reaches this location $D$, it should continue through harmlessly, and in fact notice nothing till it hits the stretched horizon which has remained at its original position.

Of course, if the stretched horizon emitted a field that reached to the location $D$, then we could argue that the incoming particle does feel something there, and nontrivial effects could be expected at that location. It would seem that the AMPS assumption of `normal physics outside the stretched horizon' would preclude any such field.  But there is in fact a field that is emitted by the matter at the stretched horizon, that does reach to the location $D$; this is just the gravitational field of the hole. We are used to ignoring this field, because we think we can change coordinates to get rid of the field. This is just the 
relation (\ref{newg}), the relation which allows us to trade a description which has the graviton propagators from the stretched horizon to the infalling particle for a description with no propagators and a new metric. The possibility of this trade is, in this context, a consequence of the principle of equivalence. 

But we have argued that this trade is exactly what fails when we reach a situation where a horizon is forming! When the gravitons in fig.\ref{fbone}(c) become dense enough to make the redshift diverge, then the graviton lines start to evolve to new configurations -- the fuzzball states -- in a theory which has such fuzzballs (fig.\ref{fbone}(d)). After this point we can no longer argue that the infalling particle will continue to fall in smoothly; rather, its energy has become converted to the energy of fuzzball states.

We do not encounter this possibility in the AMPS approach since we are not making the approximation $E\gg T$, and so the energy of the infalling particle might as well be take as zero. If $E$ can be ignored, then  we lose the above effect which makes the fuzzball surface -- the stretched horizon in our case -- move out to the location $D$; the    infalling particle must first reaching the old position of the stretched horizon before the stretched horizon can respond. 

But as we have now seen, the fuzzball surface does radiate a field -- its natural gravitational field -- to locations like $D$ outside the horizon. If the infalling particle has a very small energy, the equivalence principle allows the  trade (\ref{newg}) between graviton propagators and a new metric, and the particle continues inwards past the location $D$ without change. But if the energy $E$ of the particle is large enough to cause the classical horizon to move out to the location $D$, then we are not allowed this trade; the wavefunctional instead drifts off in a new direction in superspace, the direction of fuzzball solutions.

\b

\begin{figure}[h]
\begin{center}
 \includegraphics[scale=.72] {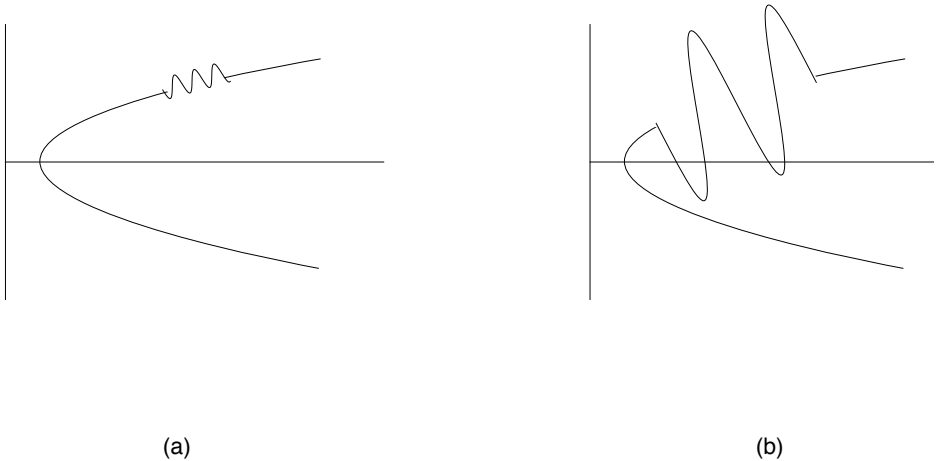}
\caption{\label{c1} (a) The fermi sea of the $c=1$ model. A small perturbation travels on locally smooth spacetime (b) A large perturbation at the same location -- one approaching the black hole threshold -- feels the bottom of the fermi sea, and does not respect the semiclassical approximation.}
\end{center}
\end{figure}

(d) For a rough analogy, we recall the $c=1$ matrix model. This can be analyzed using collective field theory \cite{dasjevicki}. In this description,   small waves on the fermi sea of eigenvalues travel like massless quanta moving at the speed of light (fig.\ref{c1}(a)). But if we take  waves of sufficiently large amplitude (corresponding to  particles with sufficiently high energy), then the deformation of the fermi sea becomes large enough to touch the bottom of the sea \cite{gross,das,pn}. In this situation the motion of the particle is altered so that it no longer behaves as a test particle on the background geometry (fig.\ref{c1}(b)). This alteration of behavior happens, in particular, at the threshold where black holes would form in the gravity theory. The moral is that we should look at the energy $E$ of the infalling particle before deciding whether something nontrivial will happen to its infall in a given geometry.

In the case of actual string theory, we should ask: what is the analogue of the fermi sea of eigenvalues that encodes the quantum structure of spacetime i the $c=1$ model? The answer is: the vast space of fuzzball solutions, which together make up a complete set of solutions at a given energy $E$. What we think of as one smooth spacetime should really  be represented as a wavefunctional on this vast superspace of solutions.  The threshold of black hole formation is the place where the mapping to a effective geometry breaks down, though it continues for a certain kinematic class of observers -- those falling in with energies $E\gg T$. 

\b

(e) Finally we address the nature of entanglement. AMPS argue that after the half-way evaporation point, the bits remaining in the hole are maximally entangled with quanta at infinity, and this prevents them from being able to  mimic free infall for an incoming particle. But if the incoming particle has a high energy $E\gg T$, then it generates {\it new} bits, which are not entangled with anything outside the hole. {\it It is the evolution of these new bits that is captured in the dual description of  fuzzball complementarity.} 

In more detail, we have the following \cite{mt2}. 
Suppose we start with a black hole of mass $M$. This hole has  $N_i=Exp[S_{bek}[M]]$ possible fuzzball states $|E_k\rangle$. Now suppose we throw in a quantum with energy $E\gg T$. The total energy after infall is  $M+E$. This implies  a  number of possible states  $N_f=Exp[S_{bek}[M+E]]$. We have
\be
{Exp[S_{bek}[M+E]]}= Exp[S_{bek}[M] + \Delta S]\approx Exp[S_{bek}[M]+{E\over T}]=Exp[S_{bek}[M]]\,  e^{E\over T}
\ee
where in the second step we have used the thermodynamic relation $dE=TdS$.
Then we find
\be
{N_f\over N_i}={Exp[S_{bek}[M+E]]\over Exp[S_{bek}[M]]}\approx  e^{E\over T}
\label{nfniq}
\ee
Using $E\gg T$ we get
\be
{N_f\over N_i}\gg 1
\label{nfni}
\ee 
This means that when we impact the fuzzball at high energy, most of the phase space allowed consists of {\it new} states that were not accessible before the impact. Since these new states are not entangled with infinity, they are not subject to the AMPS argument against a complementary description. 

\subsection{A model unitary radiation and smooth infall for $E\gg T$}

\begin{figure}[h]
\begin{center}
 \includegraphics[scale=.72] {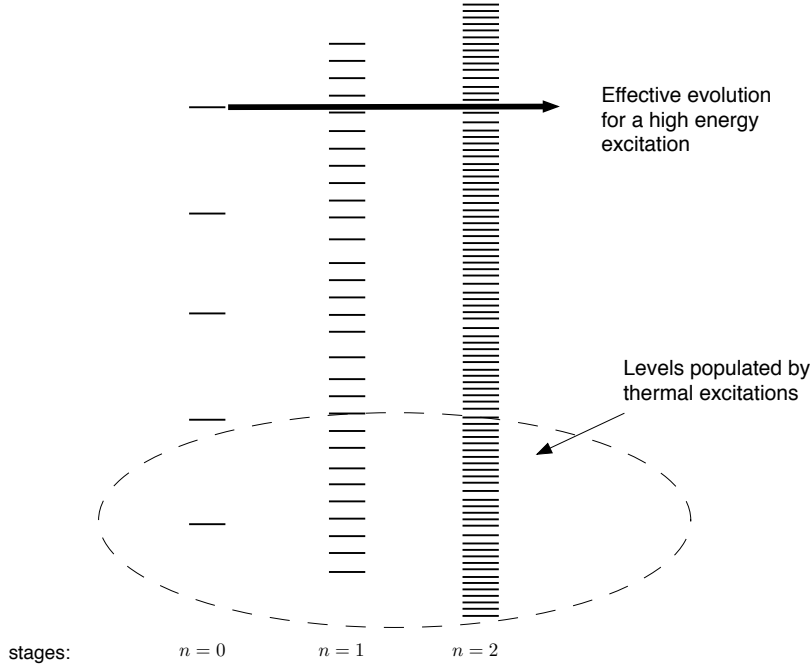}
\caption{\label{ftwof} An extension of the model in fig.\ref{ftwo}. The levels at each stage arise from a string, and the strings are coupled in the manner  (\ref{coupling2}). The strings are excited to a low temperature, but the high energy behavior of effective infall is the same as that depicted in fig.\ref{fevolution}. }
\end{center}
\end{figure}

Let us now take our above insights and make a model which:

\b

(a) Radiates energy unitarily at some low temperature $T$

\b

(b) Absorbs high energy  $E\gg T$ quanta in a way which mimics infall in an emergent coordinate direction.

\b

In the model of fig.\ref{ftwo}, we had taken an abstract set of energy levels, and noted the emergence of an effective infall direction as depicted in fig.\ref{fevolution}. We will now add to this model in two ways:

\b

(i) In fig.\ref{ftwo} we had taken only one energy level at stage $n=0$. We now let there be a band of levels at this stage as well.  We will put the entire system at a low temperature $T$, so quanta will emerge thermally from the higher stages to the stage $n=0$ just as they would in any statistical system. This process corresponds to unitary Hawking radiation from the hole.  But we also  will start with an amplitude like (\ref{qone}) in some energy level $E\gg T$ at stage $n=0$; this high energy excitation will get pulled along to higher stages $n$ just as in fig.\ref{ftwo}. This will correspond to the infall into the black hole. 

\b

(ii) In fig.\ref{ftwo} we had taken an abstract set of energy levels and assumed couplings between them. We now take a physical model which will have such couplings. We let the levels at stage $n=0$ be the vibration levels of a string of length $L_0$
\be
E^{(0)}_n={2\pi n\over L_0} +{2\pi n\over L_0}={4\pi n\over L_0}
\ee
where the two contributions come from the left and right moving parts of the wave on the string. The levels at stage $n=1$ are taken to arise for a string with length $L_1\gg L_0$
\be
E_n^{(1)}={4\pi n\over L_1}
\ee
We couple the two strings as
\be
L_{int}\sim X_0^2 X_1^2
\label{coupling2}
\ee
where $X_0, X_1$ are the excitation amplitudes of the strings at stages $n=0, 1$ respectively. This generates a coupling of the form
\be
\hat a^{(0)}_{L,p}\hat a^{(0)}_{R,p}\hat a^{\dagger(0)}_{L,q}\hat a^{\dagger(0)}_{R,q}
\ee
where a left (L) and a right (R) vibration from the string of stage $n=0$ annihilate to create a left and a right excitation on the strong at stage $n=1$. Since the level density at stage $n=1$ is much higher than that at stage $n=0$, the amplitude at each level $E_m^{(0)}$ at stage $n=0$ undergoes a fermi-golden-rule transition into the band of levels $E^{(1)}_n$ with 
\be
E^{(1)}_n\approx E_m^{(0)}
\ee
Thus even though all the levels at stage $n=1$ come from a single string, the effective behavior is quite like that of fig.\ref{ftwo} where each level at a given stage transitioned into a separate band of levels at the next stage. 

We continue in this fashion, taking a string of length $L_n\gg L_{n-1}$ for generating the levels at stage $n$, and using a coupling analogous to (\ref{coupling2}) between the strings at each stage and the next. 

\b

We now excite this set of coupled strings  at some low temperature $T$. Thus the stage $n=0$ (which in our model corresponds to the `outside' of the hole) will be populated with quanta at temperature $T$. But if we start with an excitation at stage $n=0$ at a level $E_n^{(0)}\gg T$, then we will get an evolution analogous to that in fig.\ref{fevolution}, with very little interference from the thermal excitations at temperature $T$. This gives a  model of fuzzball complementarity, depicted in fig.\ref{ftwof}.

\section{Discussion}

In this paper we have presented a model for fuzzball complementarity. The AMPS argument had ruled out the traditional complementarity proposed by Susskind \cite{susskind1}. But we have seen that fuzzball complementarity is defined somewhat differently -- it gives approximate infall into the black hole interior for infalling quanta with $E\gg T$, and is not ruled out by the AMPS argument. This difference in the two kinds of complementarities has its origins in the rather  different assumptions about how the information problem is solved in the theory. Let us review these differences. 

When complementarity was initially formulated, there was no way known to break the no-hair theorem and construct nontrivial solutions with the same quantum numbers as the black hole. All one could see was that the description of gravity was breaking down in Schwarzschild coordinates. It was the argued that somehow this breakdown of the coordinate description should lead to an effective reflection of data from the horizon for the purposes of an outside observer. But if we used good coordinates at the horizon, we would see the data fall in; thus this infalling description had to be preserved as well. Thus the idea of complementarity was that different observers see different things; there is not one wavefunction on one complete Cauchy slice; rather there are different wavefunctions seen by different observers. The reason this multiplicity of descriptions happens in the black hole and not in everyday physics is due to a basic feature of the black hole: an observer who falls inside cannot communicate with the outside. Thus we duplicate information between the inside and outside, but since this duplication cannot be checked by any observer, it is allowed \cite{susskind1}.

Thus this notion of complementarity required us to postulate  new physics. In usual quantum theory, there is one Cauchy slice and one wavefunction on this slice; it does not matter if different observers on this slice are unable to communicate with each other. For example we may take constant time slice in a flat Universe near the big crunch. Observers that are sufficiently far apart will not be able to communicate before they hit the singularity -- this is similar to the situation for observers in the black hole metric. But in normal Cosmology we do not introduce different wavefunctions for different regions of the spacelike slice. 

The problem with this proposal of complementarity was that the black hole admits just the same kind of smooth global slicing that the Cosmology does. There are no large curvatures anywhere along these slices, so it is not clear  why any new physics should kick in. The natural question becomes: if one has a complementary description where the horizon appears smooth, then won't we have the usual Hawking par production in this description, and then we will be back to Hawking's problem with growing entanglement? It would seem that complementarity cannot solve the unitarity problem.

The AMPS argument made this difficulty with traditional complementarity precise, ruling it out \cite{amps}. But now let us consider fuzzball complementarity, which is based on quite different principles. One has broken the no-hair theorem, so the surface of the hole is just like the surface of a piece of coal; there is no mystery to how information is returned from this surface. One does not need to attribute different things to different observers using different coordinate frames; the fuzzball construction is fully covariant like any other solution in gravity. Thus no new physics being invoked; all the effects we need should arise within string theory itself.

If  we want to get any kind of complementarity in this setup, we have to get it as an approximate effective description of the dynamics of the fuzzball surface. The approximation is needed because each fuzzball is different as a quantum state, and so it cannot be that all fuzzballs {\it exactly} mimic the same dynamics of infall into the traditional hole. The resulting condition $E\gg T$ then changes all the steps of the AMPS argument, and leaves the loophole for fuzzball complementarity to work.

A crucial step in this analysis was the idea of how the principle of equivalence fails at the threshold of horizon formation. AMPS \cite{amps} require that the spacetime outside the stretched horizon be normal, so an infalling particle feels nothing until it reaches the stretched horizon. But we have seen that this is not how one should pose the question; one should distinguish between the behavior felt by particles of different energies $E$. The low energy particle may indeed see nothing at a given place, but a higher energy particle would trigger fuzzball formation at the same location. This is in close analogy to what happened in the quantum gravity theory emerging from the $c=1$ matrix model. The principle of equivalence arises from a replacement of gravity propagators by a new metric as in (\ref{newg}), but it is exactly this replacement that fails when fuzzball transitions get triggered. The moral is that we must always keep in mind the wavefunctional on all of superspace; at the threshold of black hole formation, new regions of this superspace get accessed and the semiclassical approximation breaks down \cite{falling}. 

While we cannot prove at this stage that  the conjecture if fuzzball complementarity is true, we have in this paper given a model to show how it can work. It would be interesting to develop this model in more detail. Some other related ideas  can be found in \cite{martinec}.

\section*{Acknowledgements}

I would like to thank Borun Chowdhury, Per Kraus, Emil Martinec and David Turton for many helpful discussions.  This work is supported in part by DOE grant de-sc0011726.

\end{document}